\documentclass[11pt,a4paper]{article}
\usepackage{jcappub}\pdfoutput=1
\usepackage{bm}
\usepackage{amsmath}
\usepackage{hhline}
\usepackage{color}
\usepackage{xcolor}
\usepackage{slashed}
\usepackage{cancel}
\usepackage{relsize}
\usepackage[T1]{fontenc}
\usepackage{upquote}
\usepackage{changes} 
\usepackage{ulem}
\usepackage{comment}

\newcommand{\Eq}{&=&}

\newcommand{\white}[1]{{\color[rgb]{1,1,1} #1}}
\newcommand{\black}[1]{{\color[rgb]{0,0,0} #1}}

\newcommand{\hs}[1]{{\hspace{#1}}}
\newcommand{\vs}[1]{{\vspace{#1}}}
\newcommand{\mb}[1]{{\mathbf{#1}^{}}}
\newcommand{\tf}[1]{{\textsf{#1}^{}}}
\newcommand{\tx}[1]{{\text{#1}^{}}}
\newcommand{\nn}{\nonumber\\}

\newcommand{\sx}[2]{{\scalebox{#1}{#2}}}
\newcommand{\dd}{{\mathrm{d}}}
\newcommand{\OL}[1]{\overline{#1}}
\newcommand{\WB}[1]{\white{\OL{\black{#1}}}}
\newcommand{\vh}{\upsilon_h}
\newcommand{\vphi}{\upsilon_\phi}
\newcommand{\vda}{\upsilon_\delta}
\newcommand{\vD}{\upsilon_\tf{D}}

\newcommand{\mN}{m_{\hs{-0.03cm}N}}
\newcommand{\mZp}{m_{\hs{-0.03cm}Z'}}
\newcommand{\SUtwoL}{\tx{SU}(2)^{}_\tf{L}}
\newcommand{\UoneY}{\tx{U}(1)^{}_\tf{Y}}
\newcommand{\UoneD}{\tx{U}(1)^{}_\tf{D}}
\newcommand{\xiR}{\xi^{}_\tf{R}}
\newcommand{\chiL}{\chi^{}_\tf{L}}
\newcommand{\nuL}{\nu^{}_\tf{L}}
\newcommand{\gD}{g_\tf{D}}
\newcommand{\alphaD}{\alpha_\tf{D}}
\newcommand{\muD}{\mu_\tf{D}}
\newcommand{\Deltac}{\Delta^{\hs{-0.03cm}\ast}}

\begin{document}
\baselineskip=14.5pt \parskip=2.5pt

\vspace*{3em}


\title{Light Thermal Self-Interacting Dark Matter in the Shadow of Non-Standard Cosmology}

\author{
Shu$\,$-Yu\,\,Ho\footnote[1]{phyhunter@kias.re.kr},
Pyungwon\,\,Ko\footnote[2]{pko@kias.re.kr},
Dibyendu\,\,Nanda\footnote[3]{dnanda@kias.re.kr},
}

\affiliation{
Korea Institute for Advanced Study $\rm{(KIAS)}$, Seoul 02455, Republic of Korea
\vspace{3ex}}

\begin{abstract}
{In this paper, we construct a viable model for a GeV scale self-interacting dark matter (DM), where the DM was thermally produced in the early universe.\,\,Here, a new vector-like fermion with a dark charge under the $\UoneD\hs{-0.03cm}$ gauge symmetry serves as a secluded WIMP DM and it can dominantly annihilate into the light dark gauge boson and singlet scalar through the dark gauge interaction.\,\,Also, the self-interaction of DM is induced by the light dark gauge boson via the same gauge interaction.\,\,In addition to these particles, we further introduce two Weyl fermions and a doublet scalar, by which the dark gauge boson produced from $s{}^{}$-wave DM annihilations can mostly decay into active neutrinos after the dark symmetry breaking such that the CMB bound on the DM with low masses can be eluded.\,\,In order to have a common parameter region to explain the observed relic abundance and self-interaction of DM, we also study this model in a non-standard cosmological evolution, where the cosmic expansion driven by a new field species is faster than the standard radiation-dominated universe during the freeze-out of DM.\,\,Reversely, one can also use the self-interacting nature of light thermal DM to examine the non-standard cosmological history of the universe.}
\end{abstract}

\maketitle

\section{Introduction}\label{sec:1}

The existence of relic dark matter (DM) in the present universe, which constitutes almost a quarter of the total energy budget of the universe, has been a perplexing fact of particle physics and cosmology for a long while.\,\,However, the microscopic properties of DM are still unrevealed to us.\,\,Many proposals have been made by extending the standard model (SM) particle content with one or more stable or long-lived particles to make up the observed relic abundance of DM, $\Omega^\tf{obs}_\tf{DM}h^2 \simeq 0.12$~\cite{Planck:2018vyg}.\,\,For several decades, weakly interacting massive particles (or WIMP) were the most popular DM scenario to reproduce the correct relic density through the thermal freeze-out process~\cite{Feng:2010gw,
Roszkowski:2017nbc,Schumann:2019eaa,Lin:2019uvt,Arcadi:2017kky}.\,\,Nonetheless, the lack of conclusive experimental evidence inspires the exploration of different alternatives for DM beyond the WIMP paradigm.

Self-interacting DM (SIDM) has appeared as a proxy for the standard WIMP hypothesis.\,\,A 
GeV scale SIDM with a light mediator is a promising scenario since it can alleviate the small-scale issues such as too-big-to-fail, missing satellite, and core-cusp problems~\cite{Spergel:1999mh,
Tulin:2017ara,Bullock:2017xww} faced by the typical WIMP scenarios while remaining consistent with the latter at large scales, as suggested by different astrophysical observations (for recent studies on SIDM see~\cite{Ullio:2016kvy,Blum:2016nrz,Kamada:2016euw,Essig:2018pzq, Sagunski:2020spe,
Colquhoun:2020adl,Eckert:2022qia,Silverman:2022bhs,Binder:2022pmf, Girmohanta:2022dog,
Girmohanta:2022izb}).\,\,The basic difference between these two scenarios is that the typical WIMP is collisionless and SIDM has sizable self-interactions among themselves that can be parametrized in terms of the cross section to mass ratio as $\sigma^{}_\tf{SI}/m^{}_\tf{DM} \sim{\cal O}(1)\,\tx{cm}^2/\tx{g}$~\cite{Buckley:2009in,Feng:2009hw,Feng:2009mn,Loeb:2010gj,
Vogelsberger:2012ku}.\,\,Such interaction strength can be easily achieved with a GeV order DM mediated by a MeV scale particle as the self-interacting cross section can be enhanced by the mass ratio of the DM to the mediator, where $\sigma^{}_\tf{SI}/m^{}_\tf{DM} \propto m^{}_\tf{DM}/m_\tf{mediator}^4 \propto (1/m_\tf{DM}^3)(m^{}_\tf{DM}/m^{}_\tf{mediator})^4$ in the small velocity limit.

The interest in the light thermal DM paradigm ($m^{}_\tf{DM} \lesssim {\cal O}(10)\,\tx{GeV}$) has drawn more attention recently.\,\,However, in the minimal SIDM setup mentioned above, DM particles dominantly annihilate into the light mediator particles.\,\,On the other hand, the same interaction is also responsible for generating the self-interaction strength required to explain the small-scale issues that lead to the under-abundant DM relic density in the present universe.\,\,One straightforward strategy to fill the deficit in the abundance of DM can be the out-of-equilibrium decay of some long-lived particles after DM freezes out from the thermal bath~\cite{Borah:2021pet}.\,\,Recently, another mechanism has been proposed in~\cite{Borah:2022ask} where the DM relic density was compensated by the annihilation of another particle having a tiny mass splitting with  DM. However, DM models in that typical mass range face strong constraints from the cosmic microwave background (CMB) observation.\,\,According to the latest CMB measurement~\cite{Planck:2018vyg}, the DM-pair annihilation rate $\langle \sigma^{}_\tf{ann} v^{}_\tf{rel} \rangle$ during the recombination epoch has to satisfy the following condition
\begin{eqnarray}
\langle \sigma^{}_\tf{ann} v^{}_\tf{rel} \rangle
\leqslant
\frac{4.1 \times 10^{-28}\,\tx{cm}^3\,\tx{sec}^{-1}}{f^{}_\tf{eff}}
\bigg(\frac{m^{}_\tf{DM}}{\tx{GeV}}\bigg)
~,
\end{eqnarray}
where $f^{}_\tf{eff}$ is known as the efficiency factor and the specific values can be found in Refs.\,\cite{Slatyer:2015jla,Galli:2011rz,Giesen:2012rp}. However, different proposals, such as $p{}^{}{}^{}$-wave suppression of the annihilation cross section~\cite{Griest:1990kh}, asymmetry DM~\cite{Kaplan:2009ag,Iminniyaz:2011yp,Graesser:2011wi,Ho:2022tbw,Chen:2023rrl}, forbidden channel~\cite{DAgnolo:2015ujb,DAgnolo:2020mpt,Herms:2022nhd}, or $3 \to 2$ process~\cite{Hochberg:2014dra}, exist in the literature where this CMB bound can be bypassed.\,\,For well-known caveats of CMB constraints, one can see Refs.\,\cite{DAgnolo:2015ujb,
DAgnolo:2020mpt,Herms:2022nhd}.\,\,Intriguingly, the CMB bound is also pertinent for secluded DM models where DM particles annihilate into the dark sector particles and then those light particles dominantly decay into visible SM particles~\cite{Profumo:2017obk}.

However, the CMB bound is applicable only when DM (or the mediators in the case of the secluded dark sector) annihilate (decay) into electromagnetically charged particles, which can be eluded or relaxed if DM annihilates only into the electromagnetically neutral particles such as neutrinos.\,\,Nevertheless, constructing a model that can explain the correct relic density for a thermal self-interacting DM in the ballpark of a few GeV and also evades the CMB bound is a challenging mission.\,\,Here, we discuss one such possibility to find a common solution for DM self-interaction and production of correct relic density in the same ballpark.

In the conventional construction of WIMP/SIDM scenarios, the production of thermal DM is usually formulated by considering a standard radiation-dominated era right after the reheating.\,\,The timescale of this period spans from the reheating temperature $T^{}_\tf{rh}$, which depends on the inflationary scenarios, to the temperature of matter-radiation equality $T^{}_\tf{eq} \sim \tx{eV}$.\,\,It has been convincingly known that the universe was radiation-dominated about the time of the Big Bang Nucleosynthesis (BBN)~\cite{Kawasaki:2000en,Ichikawa:2005vw}.\,\,However, one cannot prohibit the possibility of some new field species other than radiation controlling the total energy budget of the universe way earlier than the BBN era.\,\,Such non-standard alternative of the usual radiation dominated universe has been studied in many different context in the literature~\cite{Chung:1998rq,Moroi:1999zb,Giudice:2000ex,Allahverdi:2002nb,Allahverdi:2002pu,
Acharya:2009zt,Allahverdi:2010xz,Monteux:2015qqa,Davoudiasl:2015vba,Berlin:2016vnh,
Tenkanen:2016jic,Berlin:2016gtr,DEramo:2017gpl,DEramo:2017ecx,Bernal:2019uqr,
Allahverdi:2020bys,Ghosh:2021wrk,Biondini:2023hek}.\,\,In this article, we will consider the evolution of thermal SIDM in the presence of a non-standard cosmological expansion history of the universe.\,\,We assume that the total energy density of the universe is governed by a new field species $\zeta$ whose energy density redshifts with the cosmic scale factor $a$ as $\rho^{}_\zeta \propto a^{-(4+n)}$ for some integers $n$.\,\,Apparently, the energy density of $\zeta$ descends faster than that of the radiation $(a^{-4})$ if $n > 0$\footnote{Various models that can produce such fast expanding universe is studied in \cite{Scherrer:2022nnz} and suggests that the scenarios with $n>2$ are difficult to construct.}.\,\,Due to such a faster redshift, $\zeta$ can be a subdominant component after the BBN epoch to be consistent with the observations~\cite{DEramo:2017gpl}.\,\,As a consequence of early $\zeta$ domination, the Hubble parameter at any given temperature becomes larger than the corresponding expansion rate for the standard cosmology at the same temperature.\,\,Faster expansion of the universe can cause the earlier decoupling of different cosmological phenomena such as the production of DM particles via freeze-out~\cite{DEramo:2017gpl} or freeze-in~\cite{DEramo:2017ecx}, and kinetic decoupling of dark radiation~\cite{Biswas:2022fga}, etc.\,which mainly depends on the competition between the interaction rate and the expansion rate of the universe.

We will show in the following sections that in such a fast-expanding universe it is possible to find a common parameter region where both relic abundance, as well as self-interactions of DM, can be satisfied.\,\,We have also proposed a UV complete scenario that is capable of evading the stringent CMB constraints.\,\,In particular, we have assumed a secluded dark sector including one fermionic DM particle which dominantly annihilates into light vector mediators.\,\,Based on the construction of the model, the light vector mediators can largely decay into neutrinos, thus not affecting the CMB data.\,\,Being a doublet under $\SUtwoL$ gauge symmetry, active neutrinos cannot directly interact with the light vector mediator and it can be generated by introducing a mixing with another chiral fermion which is charged under $\UoneD$ gauge symmetry as will be discussed in the following section.

The organization of this paper is as follows.\,\,In the next section, we first discuss the nature of the fast expansion universe and its consequences.\,\,In Sec.\,\ref{sec:3}, we briefly write down our model setup including the particle content with their charge assignment and masses and all relevant interactions.\,\,In Sec.\,\ref{sec:4}, we display the Boltzmann equation evaluating the DM relic abundance and the formula for the self-interacting cross section of DM.\,\,In Sec.\,\ref{sec:4}, we show our numerical results for the light SIDM scenario in the fast-expanding universe.\,\,The last section is devoted to discussion and conclusions.


\section{Fast expanding universe}\label{sec:2}

Before we present the detailed content of the model, let us briefly discuss the possibility of the 
non-standard cosmological expansion history and its consequences on the present study.\,\,We will consider a scenario where the early universe was not always dominated by radiation species, but rather by a species $\zeta$ whose energy density $\rho^{}_\zeta$ redshifts faster than radiation as 
\begin{eqnarray}\label{rhozeta}
\rho^{}_\zeta\big[{}^{}T(a)\big] 
\propto
a^{-(4+n)}
~,
\end{eqnarray}
where $T$ is the plasma temperature, $a$ is the cosmic scale factor, and $n > 0$.\,\,Such non-standard cosmology scenarios have been discussed in many different contexts of the literature~\cite{Turner:1983he,Ratra:1987rm,Joyce:1996cp,Copeland:1997et}.\,\,Here we assume that $\zeta$ couples feebly to the SM sector so that the only effect it will have is through the expansion rate of the universe.\,\,The realization of such a non-standard field have studied in literature by several authors~\cite{Ferreira:1997hj,Turner:1983he,Ferreira:1997au,Kallosh:2013hoa,
Kallosh:2013yoa,Joyce:1996cp,Wands:1993zm,Ratra:1987rm,Copeland:1997et,Caldwell:1997ii,
Wetterich:1994bg,Sahni:1999gb}.\,\,For example, the similar effect can be realized if the potential energy of $\zeta$ dominates than the kinetic energy and $\zeta$ is oscillating about the minimum of a positive potential and has been studied in different contexts~\cite{Kallosh:2013hoa,
Kallosh:2013yoa}.\,\,On the other hand, the scenario where the energy density of the universe was dominated by the kinetic energy of the scalar field $\zeta$ instead of the potential energy, which is often quoted as kination, has been studied in~\cite{Joyce:1996cp,Wands:1993zm,Ratra:1987rm,
Copeland:1997et}.\,\,Such theories with $n=2$ are the realizations of quintessence fluids which are also motivated to explain the accelerated expansion of the universe~\cite{Caldwell:1997ii,
Wetterich:1994bg,Sahni:1999gb}.\,\,However, here we are interested in studying the effect of such faster expansion in the relic density of SIDM.\,\,By incorporating the conservation of the total entropy $S$ in a comoving volume of the universe after the reheating epoch, $S = s{}^{}a^3 = \tx{constant}$, one can find the temperature evolution of $\rho^{}_\zeta$.\,\,The entropy density $s$ after the reheating can be expressed as
\begin{eqnarray}
s(T) \,=\, \frac{2{}^{}\pi^2}{45} {}^{} g^{}_s(T) {}^{} T^3
~,
\end{eqnarray}
where $g^{}_s(T)$ is the effective entropic degrees of freedom of the SM thermal plasma~\cite{Saikawa:2018rcs}.\,\,To be consistent with the observations of BBN, the universe has to be radiation-dominated at the time of the BBN.\,\,Let us now define a temperature ${}^{}T_r$ at which the ${}^{}\rho^{}_\zeta$ becomes equal to the radiation energy density $\rho^{}_\gamma$ below which the $\rho^{}_\gamma$ dominates the total energy budget of the universe.\footnote{Do not confuse $T_r\!$ with the reheating temperature $T_\tf{rh}$.}\,\,From Eq.\,\eqref{rhozeta}, one can write
\begin{eqnarray}
\rho^{}_\zeta(T) 
\,=\,
\rho^{}_\zeta(T_r) 
\bigg[
\frac{a(T_r)}{a(T)}
\bigg]
^{\hs{-0.03cm}4+n}
\,=\,
\rho^{}_\zeta(T_r) 
\bigg[\,
\frac{g^{}_s(T)}{g^{}_s(T_r)}
\bigg]^{\hs{-0.05cm}(4+n)/3}
\bigg(
\frac{T}{T_r}
\bigg)^{\hs{-0.12cm}4+n}
~,
\end{eqnarray}
where the last line has been derived by using the entropy conservation condition at temperature
$T_r$ and any given temperature $T$.\,\,It follows that the total energy density of the universe is~\cite{DEramo:2017gpl} 
\begin{eqnarray}
\rho^{}_\tf{tot}(T)
\,=\,
\rho^{}_\gamma(T) +
\rho^{}_\zeta(T) 
\,=\,
\rho^{}_\gamma(T) 
\Bigg\{
1 +
\frac{g^{}_\rho(T_r)}{g^{}_\rho(T)}
\bigg[\,
\frac{g^{}_s(T)}{g^{}_s(T_r)}
\bigg]^{\hs{-0.05cm}(4+n)/3}
\bigg(
\frac{T}{T_r}
\bigg)^{\hs{-0.12cm}n}
\Bigg\}
\end{eqnarray}
by setting $\rho^{}_\zeta(T_r) = \rho^{}_\gamma(T_r)$ as mentioned above, where the radiation energy density is given by~\cite{DEramo:2017gpl}
\begin{eqnarray}
\rho^{}_\gamma(T)
\,=\,
\frac{\pi^2}{30} {}^{} g^{}_\rho(T) {}^{} T^4
\end{eqnarray}
with $g^{}_\rho(T)$ is the effective energy degrees of freedom of the SM thermal plasma~\cite{Saikawa:2018rcs}.\,\,The Hubble expansion rate ${\cal H}$ of the universe at any given temperature then can be written as
\begin{eqnarray}
{\cal H}(T) 
\,=\,
\frac{1}{m^{}_\tf{Pl}}
\sqrt{\frac{8{}^{}\pi \rho^{}_\tf{tot}(T)}{3}}
~,
\end{eqnarray}
where $m^{}_\tf{Pl} = 2.435 \times 10^{18}\,\tx{GeV}$ is known as the reduced Planck mass.

At $T > T_r$, as the total energy density was dominated by $\zeta$, ${\cal H}(T)$ can be
controlled entirely by two parameters $n$ and $T_r$.\,\,However, one should be extremely cautious of the inherent impact of the presence of $\zeta$ in the total energy content and its effect on the abundance of light elements predicted by the BBN.\,\,To ensure that, $T_r$ cannot be very close to the temperatures around the BBN, where $T^{}_\tf{BBN} \sim 1\,\tx{MeV}$, and we need to follow the condition $T_r \gtrsim (15.4)^{1/n}\,\tx{MeV}$~\cite{DEramo:2017gpl}.\footnote{Since $n \geq 2$ typically, the conservative lower bound of $T_r$ is $T_r \gtrsim 4\,\tx{MeV}$.}\,\,It has been demonstrated that, if the thermal production of DM takes place before the BBN era when the expansion rate of the universe was faster than the usual radiation-dominated universe, the chemical freeze-out of DM from the thermal bath happens at an earlier time.\,\,As a consequence, the relic density of DM was overproduced.\,\,We will use this fact to enhance the relic abundance of light thermal self-interacting DM in our study.


\section{The Model Setup}\label{sec:3}

Let us now discuss the basic setup of the model.\,\,First, we enlarge the SM gauge symmetry $\tx{SU}(3)^{}_\tf{C} \otimes \SUtwoL \otimes \UoneY$ by adding a dark Abelian gauge symmetry $\UoneD$.\,\,Second, the field content of the SM is extended by introducing one vector-like fermion, $N$, two chiral fermions, $({}^{}{}^{}\xiR{}^{},\chiL)$, one complex SM singlet scalar, $\Delta$, one doublet scalar, $\Phi$, and one massive dark gauge boson, $Z'\,$(which associates with the $\UoneD$ gauge symmetry), and these new particles transform non-trivially under the $\UoneD$ gauge symmetry.\,\,The charge assignment of the SM fields and the new particles is summarized in Tab.\,\ref{tab:1}.\,\,This model belongs to a class of 2HDM plus a singlet scalar with extra 
$\tx{U}(1)^{}_\tf{H}$, that was proposed in~\cite{Ko:2012hd} as a new resolution of Higgs-mediated FCNC problem.\,\,The Yukawa sector is similar to Type-I 2HDM.\,\,Notice that the charges of these particles under the $\UoneD$ dark gauge symmetry have been chosen in such a way that $N$ becomes stable and behaves as a DM candidate thermally produced in the early universe.\,\,For simplicity, hereafter we call this model the LSIDM (Light Self-Interacting Dark Matter) model.

\begin{table}[t!]
\begin{center}
\def\arraystretch{1.3}
\begin{tabular}{|c|c|c|c||c|c|c|c|c|c|}
\hline                      
& ~$L$~ & ~$E$~ & ~$H$~ & ~$N$~ & ~\,$\xiR$~ & ~\,$\chiL$~ & ~$\Delta$~ & ~$\Phi$~ & ~\,$Z'$~ 
\\\hline\hline 
~\,$\SUtwoL$~         
& ~$\mb{2}$~ & ~$\mb{1}$~ & ~$\mb{2}$~ & ~$\mb{1}$~ & ~$\mb{1}$~ & $\mb{1}$ & $\mb{1}$ 
& ~$\mb{2}$~ & ~$\mb{1}$~
\\\hline
~\,$\UoneY$~      
& ~$-{}^{}1/2$~ & ~$-{}^{}1$~ & ~$+{}^{}1/2$~ & ~$0$~ & ~$0$~ & ~$0$~ & ~$0$~ & 
~$+{}^{}1/2$~ & ~$0$~ 
\\\hline
~\,$\UoneD$~       
& ~$0$~ & ~$0$~ & ~$0$~ & ~$+{}^{}1/2$~ & ~$+{}^{}1$~ & ~$+{}^{}1$~ & ~$+{}^{}1$~ & ~$+{}^{}1$~ & ~$0$~  
\\\hline
spin
& ~$1/2$~ & ~$1/2$~ & ~$0$~ & ~$1/2$~ & ~$1/2$~ & ~$1/2$~ & ~$0$~ & ~$0$~ & ~$1$~  
\\\hline  
\end{tabular}
\caption{Charge assignment of the fermions and scalars in the LSIDM model, where 
$L = (\,\nu \,\,\, \ell^- \,){}^\tf{T}$ is the SM lepton doublet with $\nu$ and $\ell^-$ being the SM neutrino and charged leptons, respectively, $E$ is the SM lepton singlet, and $H$ is the SM Higgs doublet.}
\vs{-0.5cm}
\label{tab:1}
\end{center}
\end{table}

In the LSIDM model, the $N$ plays the role of a secluded WIMP DM particle~\cite{Pospelov:2007mp} that mainly pair-annihilates into the $Z'$ and $\Delta$ via the dark gauge coupling.\,\,On the other hand, the $Z'$ is also a light mediator responsible for the DM self-interaction in this model.\,\,We will show the Feynman diagrams for these processes in the later section.\,\,In particular, the $Z'$ from the DM annihilation would decay into the SM-charged leptons via the kinetic mixing between the ordinary photon and the dark gauge boson after the electroweak symmetry breaking.\,\,Due to the CMB observations,\,\,this may potentially exclude the DM mass region below ${\cal O}(10)\,\tx{GeV}$ as the annihilation cross section of DM is $s{}^{}$-wave in this model~\cite{Profumo:2017obk}.\,\,However, as we shall see soon, the $Z'$ can dominantly decay into the SM neutrinos to elude the CMB constraint.\footnote{However, even in such a case, one needs to be careful about the lifetime of the $Z'$.\,\,For significantly late decay of $Z'$ to neutrinos, the amount of $\Delta{N^{}_\tf{eff}}$ would increase which can alter the prediction of BBN~\cite{Escudero:2019gzq}.}\,\,In fact, the mere reason of introducing the new scalar doublet $\Phi$ is to generate this decay process.\,\,Once the ${}^{}\Phi$ acquires the vacuum expectation value (VEV), the $Z'$ can interact with the active neutrinos through the mixing of the ${}^{}\xiR\!$ and $\nu$ as shown by the first term inside the square bracket in \eqref{Lagrangian} which was otherwise not possible with the SM gauge singlet $\Delta$. With the appearance of the right-handed fermion ${}^{}\xiR\!$ charged under the $\UoneD$ gauge symmetry, a left-handed fermion $\chiL\!$ with the same charges of ${}^{}\xiR\!$ is then introduced in order to cancel the gauge anomalies.\,\,Finally, the $\Delta$ and $\Phi$ particles trigger the $\UoneD$ symmetry breaking when they develop VEVs, after which the new particles in this model gain their physical masses.

Based on the above construction, the renormalizable Lagrangian density of the LSIDM model before the electroweak and dark symmetry breakings is written as
\begin{eqnarray}\label{Lagrangian}
{\cal L}
\Eq
({\cal D}^\kappa \hs{-0.05cm} H)^{\hs{-0.03cm}\dagger} 
({\cal D}_{\hs{-0.03cm}\kappa} H{}^{}) 
+
({\cal D}^\kappa \Phi)^{\hs{-0.03cm}\dagger} 
({\cal D}_{\hs{-0.03cm}\kappa}{}^{}\Phi) 
+
({\cal D}^\kappa \hs{-0.03cm} \Delta)^\ast 
({\cal D}_{\hs{-0.03cm}\kappa}{}^{}\Delta{}^{}) 
- 
{\cal V}(H, \Phi, \Delta)
\nn[0.2cm]
&&
{+}\,
\OL{N} \big({}^{}{}^{} i \gamma^\kappa {\cal D}_{\hs{-0.03cm}\kappa} - \mN^{} \big) N 
+
\OL{\xiR} {}^{}{}^{} 
i \gamma^\kappa {\cal D}_{\hs{-0.03cm}\kappa} {}^{}{}^{} 
\WB{\xiR}
+
\OL{\chiL} {}^{}{}^{} 
i \gamma^\kappa {\cal D}_{\hs{-0.03cm}\kappa} {}^{} 
\WB{\chiL}
- m^{}_\psi 
\sx{1.1}{\big(}\,
\OL{\xiR} {}^{} \WB{\chiL} 
+ 
\OL{\chiL} {}^{} \WB{\xiR} {}^{} 
\sx{1.1}{\big)}
\nn[0.1cm]
&&
{-}\,
\sx{1.2}{\big[}\,
y^{}_\psi {}^{} \OL{L^{}_{1\tf{L}}} \big({}^{} i \tau^2 \Phi^\ast \big) \WB{\xiR}
+
y^{}_{jk} {}^{}\OL{L^{}_{j\tf{L}}} H \WB{E^{}_{k\tf{R}}}
+ 
\mathrm{H.c.} 
\sx{1.2}{\big]}
-\frac{1}{4} C^{\kappa\omega} \hs{-0.03cm} C_{\kappa\omega}
-\frac{\epsilon}{2} B^{\kappa\omega} \hs{-0.03cm} C_{\kappa\omega}
~,
\end{eqnarray}
where ${\cal D}_{\hs{-0.03cm}\kappa} = \partial_\kappa + i g^{}_\tf{Y} {\cal Q}^{}_\tf{Y} B_\kappa + i \gD^{} {\cal Q}^{}_\tf{D} C_\kappa$ denotes the covariant derivative with $g^{}_\tf{Y}{}^{}(B_\kappa{}^{})$ and $\gD^{}{}^{}(C_\kappa)$ being the $\UoneY$ and $\UoneD$ gauge couplings (fields), respectively, and ${\cal Q}^{}_\tf{Y}\,({\cal Q}^{}_\tf{D})$ the hypercharge (dark) charge (here we omit the $\SUtwoL$ gauge fields), $B_{\kappa\omega} = \partial_\kappa B_\omega - \partial_\omega B_\kappa$, $C_{\kappa\omega} = \partial_\kappa C_\omega - \partial_\omega C_\kappa$, and $\epsilon$ the kinetic mixing parameter.\,\,For simplicity, we assume that the right-handed fermion $\xiR\!$ only couples to the first generation of the SM leptons.\,\,The $y^{}_{jk}\!$ term is nothing but the Yukawa interactions that give rise to the SM lepton masses.\,\,

The scalar potential ${\cal V}(H, \Phi, \Delta)$ is given by
\begin{eqnarray}\label{potential}
{\cal V}(H, \Phi, \Delta)
\Eq
-{}^{}{}^{}
\mu_h^2 H^{\dagger} \hs{-0.05cm} H
-
\mu_\phi^2 {}^{} \Phi^{\dagger} \hs{-0.02cm} \Phi
-
\mu_s^2 \Deltac \hs{-0.05cm} \Delta
+
\lambda^{}_h (H^{\dagger} \hs{-0.05cm} H{}^{})^2
+
\lambda^{}_\phi (\Phi^{\dagger} \hs{-0.02cm} \Phi)^2
+
\lambda^{}_\delta (\Deltac \hs{-0.05cm} \Delta)^2
\nn[0.1cm]
&&
+{}^{}{}^{}
\lambda^{}_{h\phi} (H^\dag \hs{-0.05cm} H{}^{})(\Phi^{\dagger} \hs{-0.02cm} \Phi)
+
\lambda'_{h\phi} (H^\dag \hs{-0.02cm} \Phi)(\Phi^\dag \hs{-0.05cm} H{}^{})
+
\lambda^{}_{h\delta} (H^\dag \hs{-0.05cm} H{}^{})(\Deltac \hs{-0.05cm} \Delta)
+
\lambda^{}_{\phi{}^{}\delta} (\Phi^\dag \hs{-0.02cm} \Phi)(\Deltac \hs{-0.05cm} \Delta)
\nn[0.1cm]
&&
-{}^{}{}^{}
\sqrt{2}
\sx{1.1}{\big[}\,
\muD^{} (H^\dag \hs{-0.02cm} \Phi) \Deltac + \mathrm{H.c.}
\sx{1.1}{\big]}
~.
\end{eqnarray}
Compared to the usual 2HDM, our model has no soft $Z_2-$breaking term because of $\UoneD$ symmetry.\,\, However, that term is effectively generated once $\UoneD$ symmetry is spontaneously broken and $\Delta$ develops a nonzero VEV.\,\,Note that the quadratic and quartic couplings except $\muD^{}\!$ in the scalar potential should be real if we impose the hermiticity.\,\,However, one can take $\muD^{}\!$ to be positive since the phase of $\muD^{}\!$ can be absorbed into one of the scalar fields.\,\,In the following, we will show the physical masses and relevant interactions of the particles in this model.

After spontaneously breaking of the electroweak and dark symmetries, we can decompose the scalar fields around their VEVs as
\begin{eqnarray}
H \,=\,
\begin{pmatrix}
h^+ \\[0.1cm]
\displaystyle
\frac{\vh^{} + R^{}_h + i{}^{}I^{}_h}{\sqrt{2}}
\end{pmatrix}
~,\quad
\Phi \,=\,
\begin{pmatrix}
\phi^+ \\[0.1cm]
\displaystyle
\frac{\vphi^{} + R^{}_\phi + i{}^{}I^{}_\phi}{\sqrt{2}}
\end{pmatrix}
~,\quad
\Delta \,=\, \frac{\vda^{} + R^{}_\delta + i{}^{}I^{}_\delta}{\sqrt{2}} 
~,
\end{eqnarray}
where $h^+$ and $\phi^+$ are the charged components of $H$ and $\Phi$, respectively, $\upsilon^{}_\Sigma = \langle \Sigma \rangle = \langle 0 | \Sigma | 0 \rangle$ for $\Sigma = h, \phi,\delta$ are the VEVs of the scalar fields, and $R^{}_\Sigma$ and $I^{}_\Sigma$ are their neutral real and imaginary components, respectively.\,\,Inserting these expressions into the scalar potential, the  charged and neutral components will separately mix themselves.\,\,Upon diagonalizing the mass mixing matrix of the scalar fields, the physical masses of the scalars to leading order in $\upsilon^{}_{\phi{}^{},{}^{}\delta}/\vh^{}$ are given by~\cite{Berbig:2020wve}
\begin{eqnarray}
m_h^2 \,=\, 2 {}^{} \lambda^{}_h \vh^2
~,\quad
m_{H^\pm}^2 =\, \frac{\muD^{} \vh^{}}{t^{}_\beta} - \frac{1}{2} \lambda_{h\phi}' \vh^2
~,\quad
m_{H^0}^2 =\, m_A^2 \,=\, \frac{2{}^{} \muD^{} \vh^{}}{s^{}_{2\beta}}
~,\quad
m_\delta^2 \,=\, 2 {}^{} \lambda^{}_\delta {}^{} \vD^2
~,
\end{eqnarray}
where $\vD^{} = \big(\vphi^2 + \vda^2\big){}^{1/2}$, $t^{}_\beta = \vphi^{}/\vda^{}$, $s^{}_{2\beta} = 2{}^{}t^{}_\beta/(1+t_\beta^2)$, $h$ being the SM-like Higgs boson, $H^\pm\!$ and $H^0$ are the   charged and neutral heavy CP-even bosons, respectively, $A$ is the neutral heavy CP-odd boson, and $\delta$ is the lighter neutral CP-even scalar.\,\,The mixing angles diagonalizing the scalar mass mixing matrices (expressed in terms of the VEVs $\upsilon^{}_\Sigma$ and the quadratic and quartic couplings in the scalar potential) can be found in Ref.\,\cite{Berbig:2020wve}.\,\,Throughout this work, only the light state $\delta$ will participate in our computation for the DM relic abundance.\,\,All the heavy states are unstable, successively decaying into lighter particles.

Next, we discuss the fermions in the Lagrangian density.\,\,After the dark symmetry breaking, where $\Phi$ develops the VEV, the $\xiR\!$ would mix with the active left-handed neutrino through the $y^{}_\psi\!$ term as mentioned above.\,\,Along with the $m^{}_\psi\!$ term in \eqref{Lagrangian}, the fermion mass mixing matrix in the interacting basis $\big({}^{}\nuL\,\,(\xiR)^\tf{c}\,\,\chiL{}^{}{}^{}\big)$ is given by~\cite{Farzan:2016wym,Denton:2018dqq}
\begin{eqnarray}
{\cal M}_\psi^2
\,=\,
\begin{pmatrix}
0 & y^{}_\psi \vphi^{}/\sqrt{2} & 0\,\, \\
\,y^{}_\psi \vphi^{}/\sqrt{2} & 0 & m^{}_\psi \\
0 & m^{}_\psi & 0 \\
\end{pmatrix}
~,
\end{eqnarray}
which can be diagonalized by the following matrix to leading order in $\xi^{}_\psi \equiv\, y^{}_\psi \vphi^{}/(\sqrt{2}{}^{}m^{}_\psi)$ as\footnote{A unitary matrix $\tx{diag}(1,i,1)$ is introduced in order to make the mass eigenvalues to be non-negative~\cite{Kayser:2002qs}.}
\begin{eqnarray}\label{Opsi}
{\cal O}^{}_\psi
\,=\,
\begin{pmatrix}
\,1 & 0 & 0\, \\
\,0 & i  & 0\, \\
\,0 & 0 & 1\, \\
\end{pmatrix}
\hs{-0.2cm}
\begin{pmatrix}
\,-{}^{}1& 0 & \xi^{}_\psi\,\, \\[0.1cm]
\,\xi^{}_\psi/\sqrt{2} & -{}^{}1/\sqrt{2} & 1/\sqrt{2}\, \\[0.1cm]
\,\xi^{}_\psi/\sqrt{2} & 1/\sqrt{2} & 1/\sqrt{2} \, 
\end{pmatrix}
~,\quad
\begin{pmatrix}
\nu^{}_{\tf{mL}} \\[0.05cm] (\psi^{}_\tf{R})^\tf{c} \\ \psi^{}_\tf{L}
\end{pmatrix}
\,=\,
{\cal O}^{}_\psi
\begin{pmatrix}
\nuL \\[0.02cm] (\xiR)^\tf{c} \\[-0.05cm] \chiL
\end{pmatrix}
~,
\end{eqnarray}
and the mass eigenvalues ${\cal O}^{}_\psi {\cal M}_\psi^2 {\cal O}_\psi^\tf{T} = \tx{diag}\big(0, m^{}_\psi (1+\xi_\psi^2),m^{}_\psi (1+\xi_\psi^2) \big)$.\footnote{Although the active neutrino $\nu$ mixes with the $\xiR$, however, the active neutrino keeps perturbatively massless, and its mass should originate from other mechanisms.\,\,The active neutrino mass generation is beyond the scope of this paper, one can see Ref.\,\cite{Berbig:2020wve} for possible extensions and references therein.}\,\,Notice that, due to the degenerate fermion masses in the above eigenbasis, the chiral fermions $\,\xiR$ and $\chiL\!$ are connected together to compose a Dirac fermion, $\psi$, which can be heavier than the DM mass and decay into $\nu \ell^+\ell^- $ at tree level or $\nu \gamma $ at one-loop level.

Now let us write down the gauge interactions related to the annihilation and self-interaction of DM.\,\,From the kinetic terms of $N$ and $\Delta$ in the Lagrangian density, one can easily obtain the gauge interaction of $N$ and $\Delta$ before the dark symmetry breaking as
\begin{eqnarray}\label{gauge}
{\cal L}^{}_{Z'\!N\Delta}
\,=\,
-{}^{}{}^{}\gD^{} Z'_\kappa  
\Big[
{\cal Q}^{}_N \OL{N} \gamma^\kappa \hs{-0.03cm} N 
+
i {\cal Q}^{}_\Delta
\big({}^{}
\Deltac \partial^\kappa \hs{-0.03cm} \Delta - 
\Delta {}^{}{}^{} \partial^\kappa  \hs{-0.03cm} \Deltac 
\big)
\Big]
~,
\end{eqnarray}
where $Z'$ with the mass $\mZp^{} = \gD^{} \vD^{}$ is the dark gauge boson in the mass eigenstate of the gauge bosons, and ${\cal Q}^{}_N$ and ${\cal Q}^{}_\Delta$ are the dark charges of $N$ and $\Delta$, respectively.\,\,Note that, after the dark symmetry breaking, the $Z'$ instead couples to the real singlet scalar with the gauge interaction, ${\cal L}^{}_{Z'\hs{-0.03cm}\delta} = \gD^{} \mZp^{} (Z')^2 \delta$.\,\,This means that the $\delta$ becomes unstable after the dark symmetry breaking because it can decay into a pair of $Z'$ if $m^{}_\delta > 2{}^{}\mZp^{}$.\,\,The $\delta$ will also decay after the electroweak symmetry breaking because of the Higgs portal coupling $\lambda_{h\delta}$.\,\,Finally, the gauge interaction of the active neutrino and $Z'$ can be extracted from the kinetic terms of $\,\xiR$ and $\chiL\!$ in the Lagrangian density.\,\,Using Eq.\,\eqref{Opsi}, to leading order in $\xi^{}_\psi$, we find\footnote{However, we notice that only the $\chiL\!$ has the $\nu^{}_{\tf{mL}}\!$ component due to the chirality.} 
\begin{eqnarray}
{\cal L}^{}_{Z'\!\nu} =\, 
-{}^{}{}^{}\gD^{} {}^{} \xi_\psi^2 {}^{} 
Z'_\kappa {}^{}{}^{}
\overline{\nu^{}_{\tf{mL}}} {}^{} \gamma^\kappa \WB{\nu^{}_{\tf{mL}}} ~.
\label{Zpnu}
\end{eqnarray}
As a result, the $Z'$ can mainly decay into the active neutrinos if $\xi_\psi^2 \gg \epsilon$ after the dark symmetry breaking such that the CMB constraint on the light DM mass can be evaded.\,\,We will discuss the value of the kinetic mixing parameter in the last section.


\section{Relic abundance \& Self-Interacting of DM}\label{sec:4}

\begin{figure}[t!]
\centering
\includegraphics[width=0.3\textwidth]{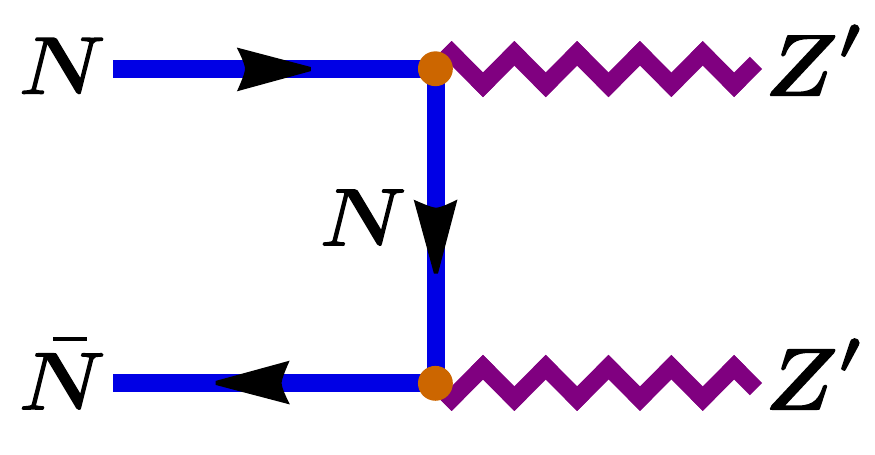}
\hs{0.2cm}
\includegraphics[width=0.3\textwidth]{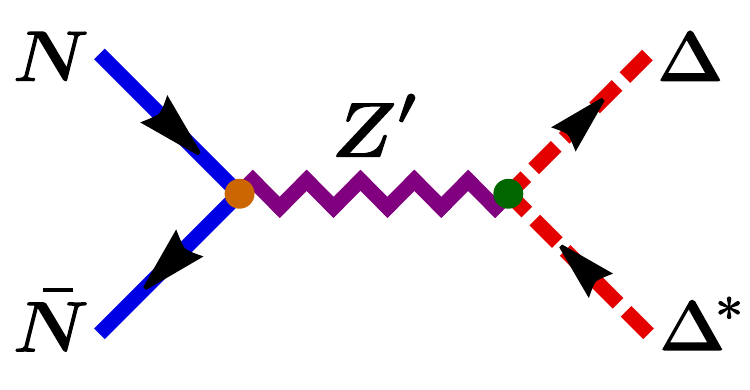}
\vs{-0.3cm}
\caption{Feynman diagrams of the annihilation processes of DM in the LSIDM model, where the arrows denote the direction of dark charge flow.\,\,The diagram crossing the final states of $N \!\bar{N} \to Z'\!Z'$ is omitted.}
\label{fig:ann}
\end{figure}

In this section, we will present the Boltzmann equation used to estimate the relic abundance of DM in a fast expanding universe, and the DM self-interaction cross section predicted by this model.\,\,Given the gauge interactions of $N$ and $\Delta$ in Eq.\,\eqref{gauge}, the Feynman diagrams of the DM annihilation processes are shown in Fig.\,\ref{fig:ann}, and the corresponding annihilation cross sections are calculated as\footnote{As it will be shown in the next section, to satisfy the observed relic abundance of DM, the $\gD^{} \sim {\cal O}(0.1)$ when $\mN^{} \sim {\cal O}(10)\,\tx{GeV}$.\,\,If we take $\mZp^{} \sim {\cal O}(10)\,\tx{MeV}$, then the energy breaking scale of the $\UoneD\!$ gauge symmetry is about $\vD^{} \sim \mZp^{}/\gD^{} \sim {\cal O}(100)\,\tx{MeV}$, which may be much lower than the DM freeze-out temperature, where $T^{}_\tf{fo} \sim \mN^{}/20 \sim 500\,\tx{MeV}$.\,\,In other words, the freeze-out of DM occurs before the dark symmetry breaking.\,\,In this case, the DM annihilation channel to the singlet scalar is $N \! \bar{N} \to \Deltac \! \Delta$ rather than $N \! \bar{N} \to \delta Z'$.\,\,Also, we have checked that the thermal masses of $\Delta$ and $Z'$ before the dark symmetry breaking are much smaller than the DM mass.\,\,Thus, they can be treated as massless particles in the DM relic abundance calculation.}
\begin{eqnarray}
\langle \sigma v \rangle^{}_{\hs{-0.05cm}N \! \bar{N} \to Z' \! Z'}
=\,
\frac{\gD^4 {\cal Q}_N^4}{16{}^{}\pi{}^{}\mN^2}
~,\quad
\langle \sigma v \rangle^{}_{\hs{-0.05cm}N \! \bar{N} \to \Deltac \! \Delta}
=\,
\frac{\gD^4 {\cal Q}_N^2 {\cal Q}_\Delta^2}{64{}^{}\pi{}^{}\mN^2}
~.
\end{eqnarray}
Assuming there is no asymmetry in DM, namely $n_{\white{{\bar{\black{N}}}}} = n_{\bar{N}}$, the Boltzmann equation of the DM number density at any given time $t$ is written as\footnote{We have numerically checked that during the freeze-out temperature of DM, the actual number densities of $\Delta$ and $Z'$ are close to their equilibrium number densities.\,\,Therefore, we have approximately
\begin{eqnarray}
\frac{\dd{}^{}n^{}_N}{\dd{}^{}t}
+
3{}^{}{\cal H}(t){}^{}n^{}_N
&\hs{-0.2cm}=\hs{-0.2cm}&
-\,
\langle {}^{} {\cal S} (\gD^{})\rangle \langle \sigma v \rangle^{}_{\hs{-0.05cm}N \! \bar{N} \to Z' \! Z'}
\Bigg[
n_N^2 - n_{Z'}^2 \frac{(n^{\tf{eq}}_N)^2}{(n^{\tf{eq}}_{Z'})^2}
\Bigg]
-
\langle {}^{} {\cal S} (\gD^{})\rangle \langle \sigma v \rangle^{}_{\hs{-0.05cm}N \! \bar{N} \to \Deltac \! \Delta}
\Bigg[
n_N^2 - n_\Delta^2 \frac{(n^{\tf{eq}}_N)^2}{(n^{\tf{eq}}_\Delta)^2}
\Bigg]
\nn
&\hs{-0.2cm}\approx\hs{-0.2cm}&
-\,
\langle {}^{} {\cal S} (\gD^{})\rangle \langle \sigma v \rangle^{}_{\hs{-0.03cm}\tf{eff}}
\Big[
n_N^2 - (n^{\tf{eq}}_N)^2
\Big]
~.
\end{eqnarray}}
\begin{eqnarray}\label{Boltzmann}
\frac{\dd{}^{}n^{}_N}{\dd{}^{}t}
+
3{}^{}{\cal H}(t){}^{}n^{}_N
\,=\,
-\,
\langle {}^{} {\cal S} (\gD^{})\rangle
\langle \sigma v \rangle^{}_{\hs{-0.03cm}\tf{eff}}
\Big[
n_N^2 - (n^{\tf{eq}}_N)^2
\Big]
~,
\end{eqnarray}
where $\langle \sigma v \rangle^{}_{\hs{-0.03cm}\tf{eff}} = \langle \sigma v \rangle^{}_{\hs{-0.05cm}N \! \bar{N} \to Z' \! Z'} + \langle \sigma v \rangle^{}_{\hs{-0.05cm}N \! \bar{N} \to \Deltac \! \Delta}$ is the effective DM annihilation cross section, $\langle {}^{} {\cal S} (\gD^{})\rangle$ is the thermally-averaged Sommerfeld enhancement factor due to the light mediator
\begin{eqnarray}
\langle {}^{} {\cal S} (\gD^{})\rangle
\,=\,
\frac{x^{3/2}}{\sqrt{4\pi}} \hs{-0.1cm}
\mathop{\mathlarger{\int}^{\infty}_0}  \dd v^{}_\tf{rel} \,
{\cal S}(\gD^{},v^{}_\tf{rel}){}^{}{}^{} v_\tf{rel}^2 {}^{}{}^{} e^{- {}^{}x{}^{}v_\tf{rel}^2/4} 
~,\quad
{\cal S}(\gD^{},v^{}_\tf{rel})
\,=\,
\frac{2{}^{}\pi \alpha^{}_\tf{D}}{v^{}_\tf{rel}}\frac{1}{1- e^{-2\pi \alpha^{}_\tf{D}/v^{}_\tf{rel}}}
\end{eqnarray}
with $x = \mN^{}/T$ being the dimensionless cosmic time and $\alpha^{}_\tf{D} = \gD^2/(4{}^{}\pi)$ the dark fine structure constant, and the number density of DM in the chemical and kinetic equilibrium with the thermal bath is
\begin{eqnarray}
n^{\tf{eq}}_N
\,=\,
\frac{g^{}_N}{2{}^{}\pi^2} {}^{} \mN^2 T K^{}_2 \hs{-0.03cm} (x)
\,\approx\,
g^{}_N 
\bigg(\frac{x}{2{}^{}\pi}{}^{}{}^{}\bigg)^{\hs{-0.12cm}3/2}
T^3 e^{-x}
\end{eqnarray}
with $g^{}_N = 2$ being the spin state number of $N$, and $K^{}_2$ denotes the modified Bessel function of the second kind.\,\,As usual, we define the comoving 
number density of DM, $Y^{}_N = n^{}_N/s$, then the Boltzmann equation assuming $g^{}_\rho(x) \simeq g^{}_s(x)$ and $\dd g^{}_\rho(x)/dx \simeq 0$ becomes
\begin{eqnarray}\label{BoltzmannY}
\frac{\dd{}^{}Y^{}_N}{\dd{}^{}x}
\,=\,
-{}^{}
\frac{s(x)}{x {\cal H}(x)}\langle {}^{} {\cal S} (\gD^{})\rangle
\langle \sigma v \rangle^{}_{\hs{-0.03cm}\tf{eff}}
\Big[
Y_N^2 - (Y^{\tf{eq}}_N)^2
\Big]
~,
\end{eqnarray}
where the equilibrium comoving number yield of DM is given by
\begin{eqnarray}
Y^{\tf{eq}}_N
\,=\,
\frac{45}{4 \pi^4}
\frac{g^{}_N}{g^{}_s(x)} {}^{}
x^2
K^{}_2 \hs{-0.03cm} (x)
\,\approx\,
\frac{45\sqrt{2}}{8{}^{}\pi^{7/2}}
\frac{g^{}_N}{g^{}_s(x)} {}^{}
x^{3/2} e^{-x}
~.
\end{eqnarray}
With an appropriate initial condition, the Boltzmann equation \eqref{Boltzmann} can be numerically solved, and the present density of DM can be estimated by using the following relation~\cite{Bhattacharya:2019mmy}
\begin{eqnarray}
\Omega^{}_\tf{DM} \hat{h}^2 
\,\simeq\,
5.49 \times 10^8 {}^{}
\sx{0.9}{\bigg(}
\frac{\mN^{}}{\text{GeV}}
\sx{0.9}{\bigg)} {}^{}
Y^{}_N(x \to \infty)
~,
\end{eqnarray}
where $\hat{h} \simeq 0.7$ is the normalized Hubble constant.\,\,Since the annihilation cross sections of DM only depend on the dark gauge coupling and the mass of DM, after imposing the observed DM relic abundance data, one can get the $\gD^{}\!$ as a function of $\mN^{}\hs{-0.05cm}$ given the values of $n$ and $T_r\hs{-0.05cm}$ in the fast-expanding universe scenario discussed in Sec.\,\ref{sec:2}.

\begin{figure}[t!]
\centering
\includegraphics[width=0.3\textwidth]{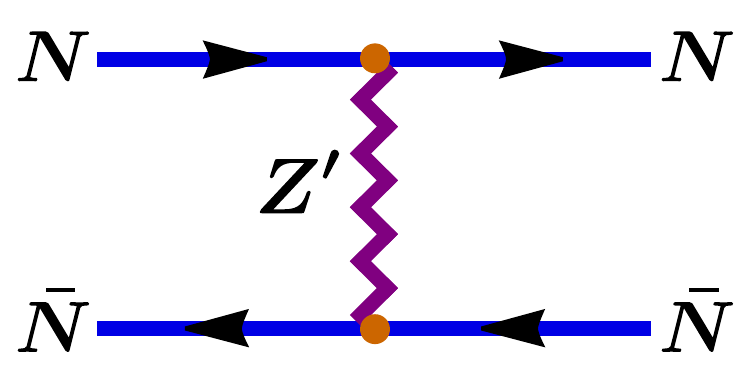}
\hs{0.2cm}
\includegraphics[width=0.3\textwidth]{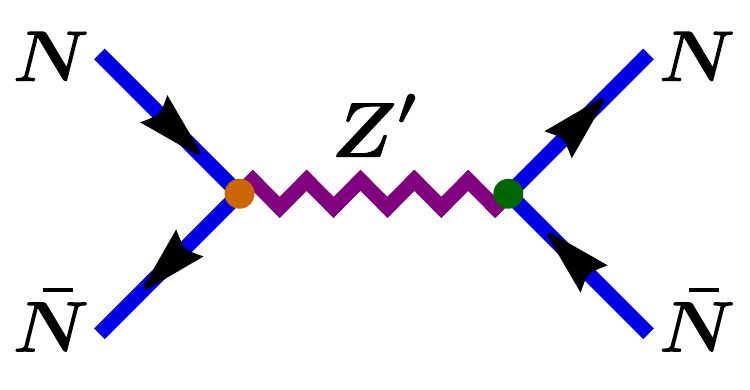}
\vs{-0.3cm}
\caption{Feynman diagrams of the self-interacting processes of DM in the LSIDM model, where the other diagrams can be obtained by flipping the dark charge flow and/or crossing the final states.}
\label{fig:self}
\end{figure}

With the same gauge interaction, the Feynman diagrams of the DM self-interacting processes are depicted in Fig.\,\ref{fig:self}.\,\,Such self-interactions mediated by MeV scale $Z'$ can be represented by a Yukawa-type potential,
\begin{eqnarray}
V(r) \,=\, \pm {}^{} \frac{\alphaD^{}}{r} {}^{} e^{- {}^{} \mZp r} ~,
\end{eqnarray}
where the $+$ or $-$ sign in the above equation dictates the nature of the potential whether repulsive or attractive.\,\,To study the relevant effects of forward scattering divergences for the generic self-interactions of DM, the transfer cross section can be defined as~\cite{Tulin:2017ara,
Feng:2009hw,Tulin:2013teo} 
\begin{eqnarray}
\sigma^{}_\tf{T} \,=\, 
\mathop{\mathlarger{\int}} \dd\Omega \, \big(1-\cos{\theta}\big) \frac{\dd \sigma}{\dd \Omega} ~.
\end{eqnarray}
Depending on the values of the relevant parameters (in particular $\gD^{},\,\mN^{},\,\mZp^{}$) in the self-interaction of DM, the transfer cross section can be categorized as follows.\,\,In the limit $\alphaD^{} \mN^{}/\mZp^{} \ll 1$, known as the Born limit, $\sigma^{}_\tf{T}$ for both attractive and repulsive potential can be expressed as
\begin{eqnarray}
\sigma^\tf{Born}_\tf{T}
\,=\,
\frac{8{}^{} \pi \alphaD^2}{\mN^2 v^4} 
\Bigg[\ln  \hs{-0.05cm} \bigg(1 + \frac{\mN^2 v^2}{\mZp^2}\bigg)
-
\frac{\mN^2 v^2}{\mZp^2 + \mN^2 v^2}
\Bigg]
~.
\end{eqnarray}
On the other hand, in the regime outside the Born limit, $\alphaD^{} \mN^{}/\mZp^{} \geq 1$, the impact of attractive or repulsive potential will have different behavior.\,\,In the already existing literature, this region has been classified into two different sub-categories.\,\,In the limit $\mN^{} {}^{}v/\mZp^{} \gg 1$, known as the classical limit, the transfer cross sections for attractive potential can be expressed by~\cite{Tulin:2013teo,Tulin:2012wi,Khrapak:2003kjw}
\begin{eqnarray}
\sigma^\tf{Classical}_\tf{T}
\,=\,
\begin{cases}
\displaystyle
\frac{4{}^{}\pi}{\mZp^2} {}^{} \beta^2 \ln \hs{-0.05cm} \bigg(1 + \frac{1}{\beta} \bigg) 
& \text{if \, $\beta\leq 10^{-1}$} 
\\[0.4cm]
\displaystyle
\frac{8{}^{}\pi}{\mZp^2} \frac{\beta^2}{1 + 1.5{}^{}\beta^{1.65}}  
& \text{if \, $10^{-1} \leq \beta \leq 10^{3}$}
\\[0.4cm]
\displaystyle
\frac{\pi}{\mZp^2} 
\bigg(1 + \ln \beta - \frac{1}{2 \ln \beta} \bigg)^{\hs{-0.13cm}2} 
& \text{if \, $\beta\geq 10^3$}
\end{cases}       
\end{eqnarray}
whereas the cross section for the repulsive potential is 
\begin{eqnarray}
\sigma^\tf{Classical}_\tf{T}
\,=\,
\begin{cases}
\displaystyle
\frac{2{}^{}\pi}{\mZp^2} \beta^2 \ln \hs{-0.05cm} \bigg(1 + \frac{1}{\beta^2} \bigg) 
& \text{if \, $\beta\leq 1$}
\\[0.4cm]
\displaystyle
\frac{\pi}{\mZp^2} \Big[ \ln \hs{-0.05cm} \big(2\beta) - \ln\ln \hs{-0.05cm} \big(2\beta) {}^{} \Big]^{\hs{-0.05cm}2} 
& \text{if  \, $\beta \geq 1$}
\end{cases}       
~,
\end{eqnarray}
where $\beta$ is defined as $\beta = 2{}^{}\alphaD^{} \mZp^{}/(\mN^{} {}^{}v^2)$.\,\,There can be another possible regime (known as the resonance regime) of interest to consider the mild velocity dependence in the transfer cross section that is needed to explain the small-scale issues of cold DM.\,\,In this regime, $\alphaD^{} \mN^{}/\mZp^{} \geq 1$ and $\mN^{} {}^{} v/\mZp^{} \lesssim 1$, there is no analytical formula for $\sigma^{}_\tf{T}$ exist in the literature.\,\,To estimate the cross section, one needs to solve the non-relativistic Schrodinger equation by using the partial wave analysis.\,\,However, by approximating the Yukawa potential to be a Hulthen potential with a screening mass $\vartheta$, 
\begin{eqnarray}
V(r)\,=\, 
\pm 
\frac{\alphaD^{} \vartheta {}^{}{}^{} e^{- {}^{} \vartheta {}^{} r}}
{1 - e^{- {}^{} \vartheta {}^{} r}}
~,
\end{eqnarray}
one can use the non-perturbative results for $s{}^{}$-wave scattering to calculate the transfer cross section in the resonance regime as~\cite{Tulin:2013teo}
\begin{eqnarray}
\sigma^\tf{Hulthen}_\tf{T}
\,=\, 
\frac{16{}^{}\pi \sin^2 \vartheta^{}_0}{\mN^2 v^2} 
~,
\end{eqnarray}
where 
\begin{eqnarray}
\displaystyle
\vartheta^{}_0 \,=\, 
\tx{arg}
\Bigg[
\frac{i}{\Gamma(\varepsilon^{}_+)\Gamma(\varepsilon^{}_-)} {}^{}
\Gamma \bigg(\frac{i {}^{} \mN^{}v}{\kappa {}^{} \mZp^{}}\bigg)
\Bigg]
\end{eqnarray}
with $\Gamma(z)$ the Gamma function, and $\kappa \approx 1.6$ a dimensionless number.\,\,The expression of $\varepsilon^{}_{\pm}$ depending on the nature of the potential whether attractive or repulsive are given by
\begin{eqnarray}
\varepsilon^{}_{\pm} 
\,=\, 
\begin{cases}
\displaystyle
1 + \frac{i{}^{}\mN^{}v}{2{}^{}\kappa{}^{}\mZp} 
\pm 
\sqrt{\frac{\alphaD^{} \mN^{}}{\kappa{}^{}\mZp^{}} - 
\frac{\mN^2 v^2}{4{}^{}\kappa^2 \mZp^2}} 
& \tx{Attractive}
\\[0.4cm]       
\displaystyle
1 + \frac{i{}^{}\mN^{}v}{2{}^{}\kappa{}^{}\mZp^{}} 
\pm 
i{}^{}\sqrt{\frac{\alphaD^{} \mN^{}}{\kappa{}^{}\mZp^{}} +
\frac{\mN^2 v^2}{4{}^{}\kappa^2 \mZp^2}} 
& \tx{Repulsive}
\end{cases}
~.
\end{eqnarray}
\begin{figure}[t!]
\centering
\includegraphics[width=0.6\textwidth]{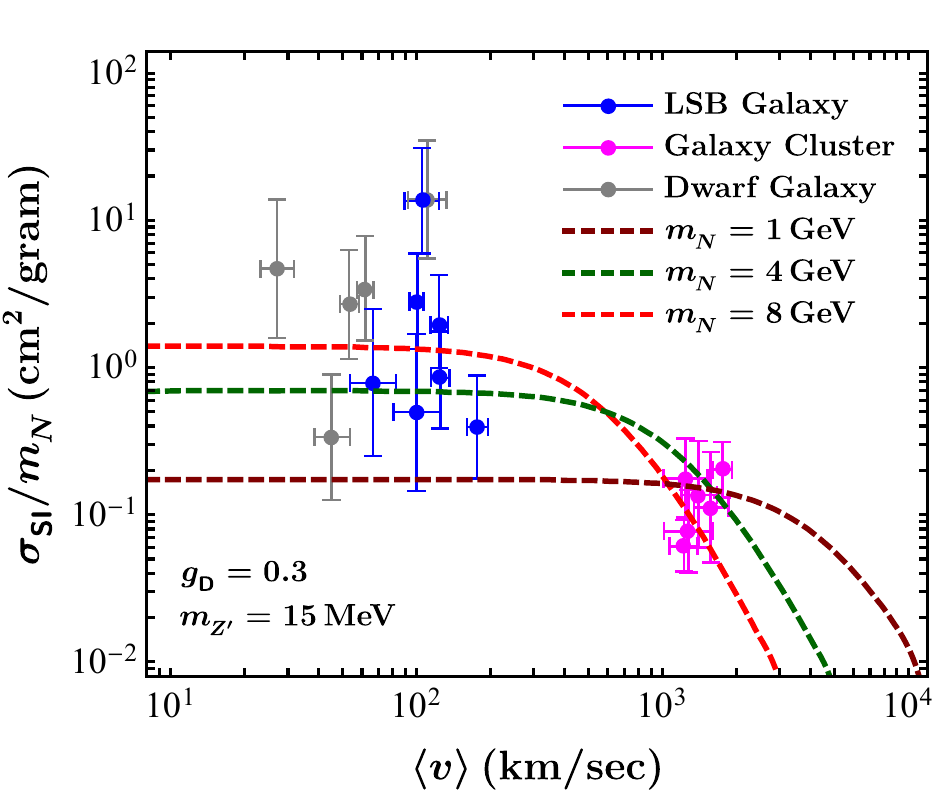}
\caption{The self-interaction cross section per unit mass of DM is shown as a function of the average collision velocity for three benchmark values of DM mass. The observational data points with their error bars are shown by blue, magenta, and grey points as in~\cite{Kaplinghat:2015aga}.}
\label{fig:SI_vel}
\end{figure}

We show the dependence of velocity on the self-interaction cross section per unit mass of DM in Fig.\,\ref{fig:SI_vel} where model dependence has been shown for three different DM masses.\,\,The other parameters are fixed as $\gD^{}=0.3$ and $\mZp^{} =15\,\tx{MeV}$.\,\,The observational data points with their error bars from low surface brightness (LSB) galaxies, galaxy clusters, and dwarf galaxies are shown by blue, magenta, and grey points respectively as in~\cite{Kaplinghat:2015aga}.\,\,It is important to note that the model we consider here can describe the observed velocity dependence of the self-interacting cross section significantly well.


\section{Constraints on the scalar sector parameters}
\begin{figure}[t!]
\centering
\includegraphics[scale=0.46]{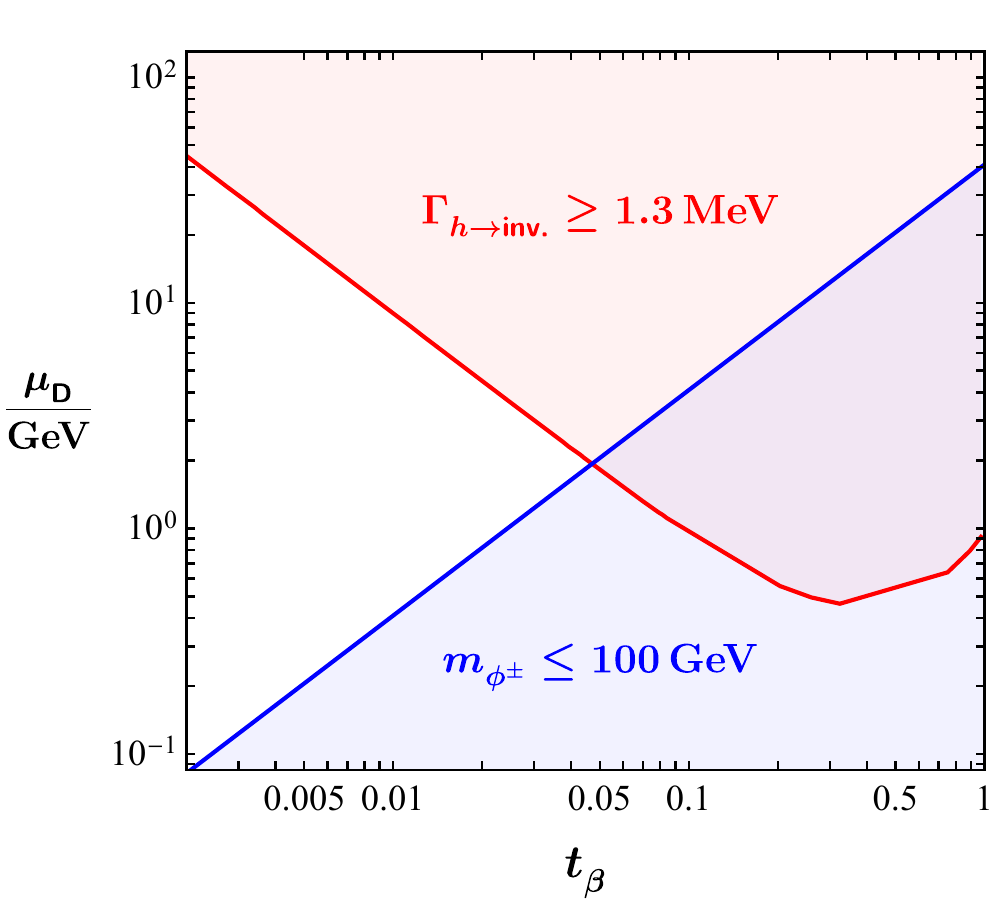}
\vs{-0.3cm}
\caption{Available parameter space in $(t^{}_\beta,\muD^{})$ plane from $\Gamma_{h \rightarrow \tf{inv.}}$ limit and the mass of the charged scalar.\,\,We have fixed the other parameters as $\lambda^{}_{h\delta} = 10^{-3}$ and $\lambda_{h\phi} + \lambda_{h\phi}^\prime = 0.005$.}
\label{fig:Hinv}
\end{figure}

Due to the presence of the new scalar doublet in the particle spectrum, the electroweak precision observables will change than the SM prediction.\,\,The expressions for the electroweak precision observables for extended scalar sectors have been derived in~\cite{Grimus:2007if,Grimus:2008nb}. The limits on $\mathbb{S}$ and $\mathbb{T}$ parameters have been greatly improved and can be expressed as $\mathbb{S} = -0.01 \pm 0.07$ and $\mathbb{T} = 0.04 \pm 0.06$~\cite{ParticleDataGroup:2022pth}.\,\,However, as we discussed earlier, the components of the doublet scalar can be heavy and stay in the allowed for $\big| m_{\phi^\pm} - m_{H^0} \big|\leq 120\,\tx{GeV}$.\,\,Along with the $\mathbb{S}$ and $\mathbb{T}$ parameters, the presence of same doublet can also enhance the $h \rightarrow \gamma \gamma$ through one-loop process.\,\,The present limit from CMS~\cite{CMS:2021kom} on $h \rightarrow \gamma \gamma$ reads as ${\cal B}(h \rightarrow \gamma \gamma)^{}_\tf{exp.}/{\cal B}(h \rightarrow \gamma \gamma)^{}_\tf{SM} = 1.12 \pm 0.09$.\,\,By using the expression given in~\cite{Djouadi:2005gi}, the model is consistent for $\lambda_{h\phi} \leq 5 \times 10^{-4}$ with $m_{\phi^\pm} \geq 100\,\tx{GeV}$.\,\,The presence of the light scalar $\Delta$ will affect the invisible decay width of Higgs due to the two new decay channels $h\rightarrow \Delta \Delta$ and $h \rightarrow Z^\prime Z^\prime$. The decay widths of these channels can be expressed as
\begin{eqnarray}
\Gamma^{}_{h \rightarrow \Delta \Delta} 
\,=\,
\Gamma^{}_{h \rightarrow Z^\prime Z^\prime} 
\,=\, 
\frac{\upsilon_h^2}{32 \pi m^{}_h} 
\bigg[ 
\lambda_{h\delta} 
\cos^2{t^{}_\beta} +
\big( \lambda_{h\phi} + \lambda_{h\phi}' \big) \sin^2{t^{}_\beta} -
\frac{\muD^{}}{\upsilon_h} \sin{\big(2{}^{}t^{}_\beta\big)}
\bigg]
~,
\end{eqnarray}
provided $t_{\beta}$ is not very small. In Fig.\,\ref{fig:Hinv}, we show the impact of the invisible decay width of Higgs and the mass of the charged scalar in $t^{}_\beta$ versus $\muD^{}$ plane which excludes a significant amount of parameter space. The red-shaded region represents the currently excluded region from Higgs invisible decay, $\Gamma_{h \rightarrow \tf{inv.}} \leq 1.3\,\tx{MeV}$ \cite{ParticleDataGroup:2022pth}, whereas the blue line corresponds to $m_{\phi^\pm} = 100$ GeV. Here, we have fixed the quartic couplungs $\lambda_{h\phi}' = 10^{-3}$ and $\lambda^{}_{h\phi} = 5\times 10^{-4}$ guided by the $h\rightarrow \gamma \gamma$ constraints as discussed above. 

As discussed earlier, the presence of $\Phi$ is extremely important for generating the mixing between ${}^{}\xiR\!$ and $\nu$ which introduces the new decay channel for $Z^\prime$ to neutrinos. However, the mixing parameter, $\xi_{\psi}$, does not depend on the masses of $\Phi$ rather it depends only on $y_{\psi},\,\upsilon_{\phi},\,\text{and } m_{\psi}$. In the upcoming section, we will discuss the numerical results of our model in detail. To ensure phenomenological consistency with all the aforementioned experimental constraints, we have assumed that the masses of all components of $\Phi$ are on the order of $\mathcal{O}(100 \text{ GeV})$. Importantly, these heavy scalars are not anticipated to influence the specific phenomenologies of interest.

\section{Numerical results}\label{sec:5}
Finally, let us discuss the implications of the faster expansion of the universe on the LSIDM model.\,\,The main motivation of this work is to find a common parameter space that can satisfy the correct relic abundance and self-interaction strength required for cold DM.\,\,Another critical issue for light thermal DM is that it faces a stringent constraint from the CMB data due to the significant amount of energy injected into the thermal plasma from the large annihilation rates of DM.\,\,However, as we have discussed in Sec.\,\ref{sec:3}, the CMB restrictions on the light DM can be escaped in our model because of the decay of $Z'$ into the active neutrinos.

\begin{figure}[t!]
\centering
\includegraphics[width=0.493\textwidth]{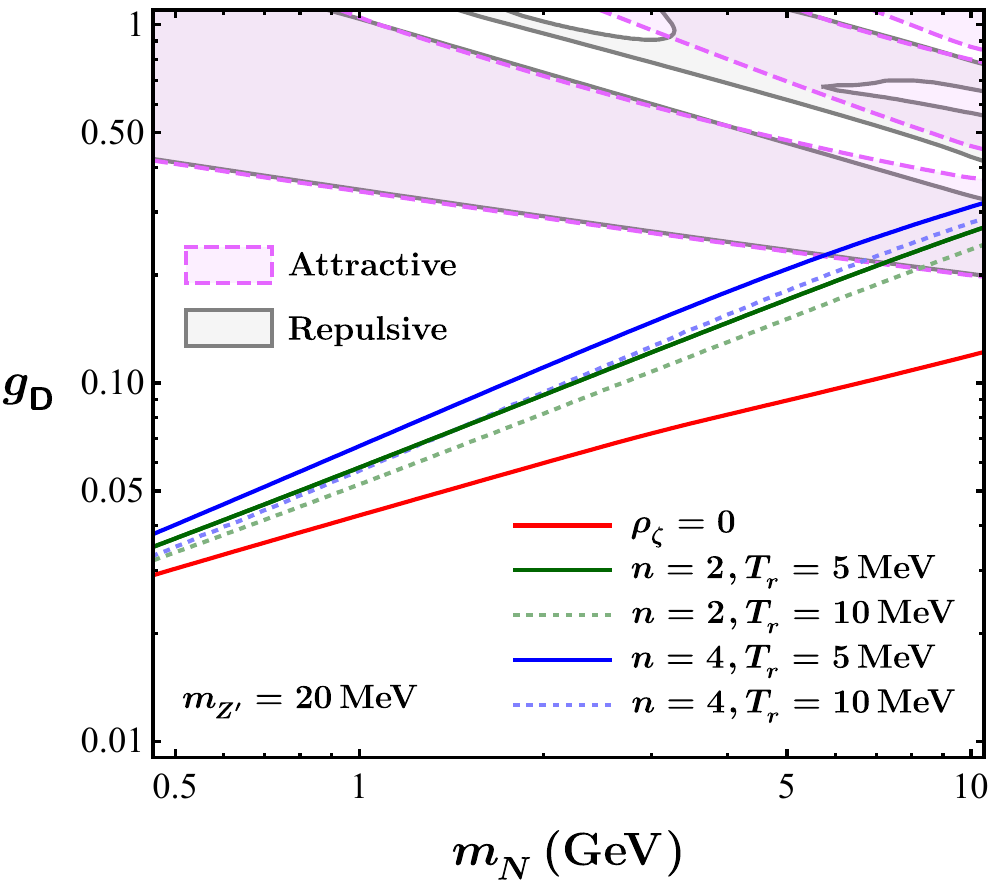}
\hs{0.0cm}
\includegraphics[width=0.493\textwidth]{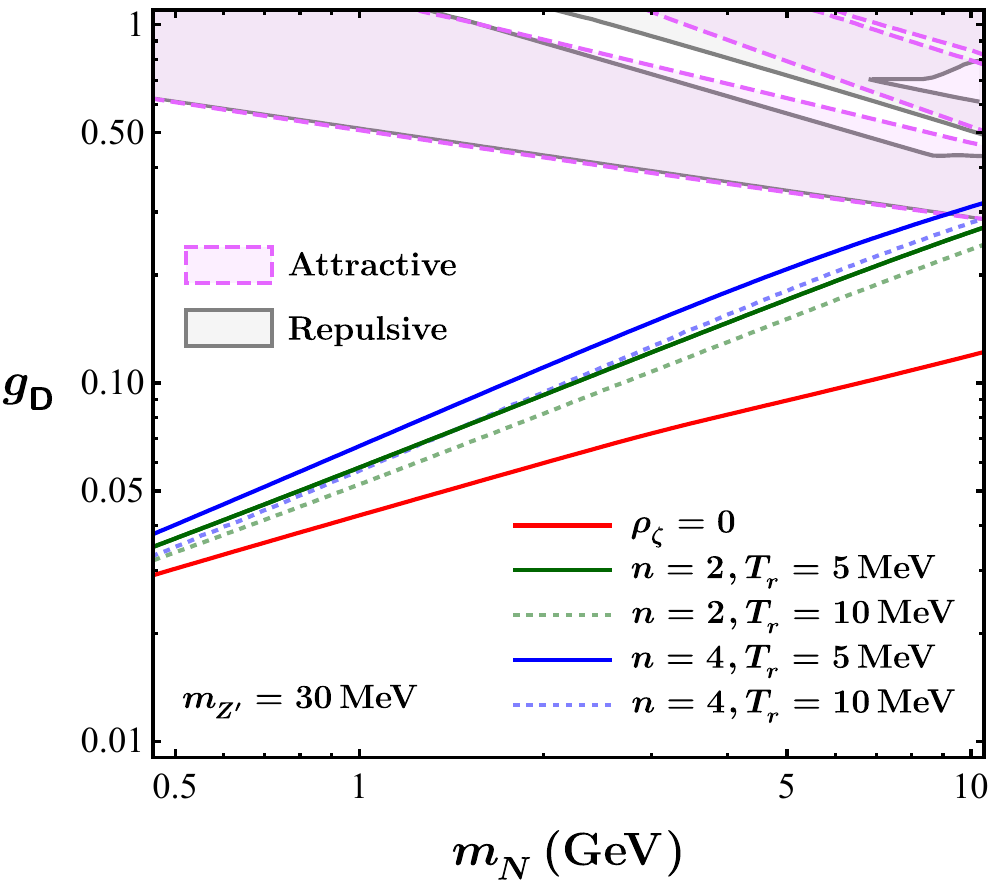}
\vs{-0.7cm}
\caption{Allowed parameter space of $\gD^{}\!-\mN^{}$ plane in the LSIDM model for two choices of $\mZp^{}$, where the left panel is for $\mZp^{} = 20\,\tx{MeV}$ and the right panel is for $\mZp^{} = 30\,\tx{MeV}$.\,\,In both figures, the light magenta and grey regions represent the required $\gD^{}\!$ and  $\mN^{}\!$ to explain the self-interactions of DM, and color lines satisfy the observed relic abundance of DM, $\Omega^\tf{obs}_\tf{DM}  \simeq 0.12$.\,\,The red line corresponds to the standard radiation-dominated universe and the green and blue lines (with solid lines for $T_r = 5\,\tx{MeV}$ and dotted lines for $T_r = 10\,\tx{MeV}$) correspond to the fast expanding universe for $n = 2$ and $n = 4$, respectively.}
\label{fig:gDvsmN}
\end{figure}

We show in Fig.\,\ref{fig:gDvsmN} the parameter space of the DM mass and the dark gauge coupling for two different values of $Z'\!$ mass, where $Z'\!$ acts as final state particles in the case of DM annihilation and as a mediator in the case of DM self-interaction.\,\,In both figures, the light magenta (light gray) region indicates the required dark gauge coupling and DM mass to explain the DM self-interacting cross section for attractive (repulsive) potential from the astrophysical observations.\,\,For both these plots, we have fixed the velocity $v \sim 200\, \tx{km/s}$ which falls in the range of observations from the LSB galaxies as shown in Fig.\,\ref{fig:SI_vel}.\,\,One can see that the region that satisfies the required self-interaction shifts towards the larger coupling if we increase $\mZp^{}$.\,\,This can be easily understood.\,\,Because when the propagator mass is heavier, the stronger coupling is needed to account for the same value of the DM self-interacting cross section.\,\,We have shown the dependence of relic density on $\gD^{}$ and $\mN^{}$ by solid lines or dotted lines.\,\,The red solid line shows the required dark gauge coupling as a function of DM mass for getting the correct DM relic abundance in the standard radiation-dominated universe where there was no $\zeta$ presented in the early universe.\,\,One can notice that in the absence of $\rho^{}_\zeta$, the demanded $\gD^{}$ to have the correct relic density of DM (the red line) is much smaller in comparison to the one needed to explain the self-interaction of DM (the light magenta and grey regions).\,\,As mentioned previously, once $\zeta$ is introduced into the energy spectrum and the expansion of the universe becomes faster than the radiation-dominated universe, the red line which satisfies the observed density of DM shifts toward the higher values of couplings.\,\,This can be understood as follows\,\,:\,\,in the fast expanding universe, for a $\gD^{}\hs{-0.05cm}$ that was giving the correct DM relic density, the DM freezes out earlier than the standard cosmology and as a result, will overclose the universe.\,\,To keep DM in the thermal bath for a longer period of time or to suppress the enhancement of freeze-out abundance of DM we need to strengthen the gauge coupling $\gD^{}$.\,\,Here we present the effect of faster expansion for four different combinations of $n$ and $T_r$, where green lines are for $n = 2$ and blue lines are for $n = 4$ (both solid and dotted).\,\,We have already explained that larger $n$ corresponds to a faster expansion rate and one needs stronger coupling strengths to satisfy the correct relic abundance of DM.\,\,As we increase the effect of $\zeta$ further the line enters more into the light magenta and grey regions.\,\,Here, one important thing to notice is that for heavier $Z'$, we need stronger dark gauge coupling $\gD^{}$ to satisfy the required self interaction.\,\,However, the relic density of DM, in the concerned mass range, is almost independent of $\mZp^{}$.\,\,As a result, for sufficiently heavier $Z'$, even faster expansion can not provide a common parameter space which can accommodate both the correct relic density as well as the required self-interaction.

\section{Discussion \& Conclusions}\label{sec:6}
Before concluding this work, let us discuss the thermalization of the DM with the SM plasma as we assume that the DM particle was produced thermally in the early universe.\,\,In order for the DM to have a common temperature with the SM thermal plasma, the energy transfer rate $\gamma^{}_N(T)$ between the DM particle and the SM particles should be bigger than the Hubble expansion rate ${\cal H}(T)$.\,\,Following Refs.\,\cite{Gondolo:2012vh,Ho:2022erb} and assuming the SM particles are massless compared to the DM, we have
\begin{eqnarray}
\gamma^{}_{N}(T) \,=\,
\frac{31{}^{}\pi^3 \gD^2 {\cal Q}_N^2 {}^{} g_e^2 c_\tf{w}^2 \epsilon^2}{189{}^{}x^6}
\frac{\mN^5}{\mZp^4}
~,
\end{eqnarray}
where $g^{}_e = (4\pi\alpha^{}_\tf{EM})^{1/2} \simeq 0.3$ with $\alpha^{}_\tf{EM}$ the fine structure constant, and $c_\tf{w} = \cos\theta_\tf{w}$ with $\theta_\tf{w}$ the weak mixing angle.\,\,Also, the Hubble expansion rate in the fast-expanding universe scenario for $T \gg T_r$ can be approximated as
\begin{eqnarray}
{\cal H}(T) 
\,\simeq\,
\sqrt{\frac{\pi^2 g^{}_\rho(T)}{90}}
\frac{T^2}{m^{}_\tf{Pl}}
\bigg(\frac{T}{T_r}\bigg)^{\hs{-0.15cm}n/2}
~.
\end{eqnarray}
Imposing the thermalization condition of the DM and the SM particles, $\gamma^{}_N(T) \gtrsim 2{}^{}{\cal H}(T)$~\cite{Gondolo:2012vh}, at the freeze-out temperature of DM, we find the lower bound of the kinetic mixing parameter as
\begin{eqnarray}
\hs{-0.2cm}
\epsilon 
\,\gtrsim\,
\begin{cases}
\displaystyle
1.31 \times 10^{-10} 
\bigg[\frac{g^{}_\rho(x^{}_\tf{fo})}{60}\bigg]^{\hs{-0.05cm}1/4}
\bigg(\frac{x^{}_\tf{fo}}{20}\bigg)^{\hs{-0.13cm}3/2}
\bigg(\frac{\gD^{}}{0.3}\bigg)^{\hs{-0.15cm}-1}
\bigg(\frac{\mN^{}}{10\,\tx{GeV}}\bigg)^{\hs{-0.15cm}-1}
\bigg(\frac{\mZp^{}}{10\,\tx{MeV}}\bigg)^{\hs{-0.13cm}2}
\bigg(\frac{T_r}{10\,\tx{MeV}}\bigg)^{\hs{-0.15cm}-1/2}&
\\
\hs{12.5cm}\tx{for} \,\,\, n\,=\,2&
\\[0.15cm]
\displaystyle
9.28 \times 10^{-10} 
\bigg[\frac{g^{}_\rho(x^{}_\tf{fo})}{60}\bigg]^{\hs{-0.05cm}1/4}
\bigg(\frac{x^{}_\tf{fo}}{20}\bigg)^{\hs{-0.13cm}3/2}
\bigg(\frac{\gD^{}}{0.3}\bigg)^{\hs{-0.15cm}-1}
\bigg(\frac{\mN^{}}{10\,\tx{GeV}}\bigg)^{\hs{-0.15cm}-1/2}
\bigg(\frac{\mZp^{}}{10\,\tx{MeV}}\bigg)^{\hs{-0.13cm}2}
\bigg(\frac{T_r}{10\,\tx{MeV}}\bigg)^{\hs{-0.15cm}-1/2}&
\\
\hs{12.5cm}\tx{for} \,\,\, n\,=\,4&
\end{cases}
\hs{-0.4cm}.
\end{eqnarray}
On the other hand, in order for the $Z'$ mainly decays into the active neutrinos, the decay rate of $Z' \to \nu\bar{\nu}$ should be bigger than that of $Z' \to e^+e^-$.\,\,Using the kinetic mixing term in Eq.\,\eqref{Lagrangian} and the interaction of the $Z'$ with the SM neutrinos in Eq.\,\eqref{Zpnu}, their decay rates are calculated as
\begin{eqnarray}
\Gamma^{}_{Z' \to e^+e^-}
\,=\,
\frac{g_e^2 c_\tf{w}^2 \epsilon^2}{12{}^{}\pi} \mZp^{}
~,\quad
\Gamma^{}_{Z' \to \nu\bar{\nu}}
\,=\,
\frac{\gD^2 \xi_\psi^4}{24{}^{}\pi} \mZp^{}
~.
\end{eqnarray}
Requiring that $\Gamma^{}_{Z' \to \nu\bar{\nu}} \gg \Gamma^{}_{Z' \to e^+e^-}$, we find the upper limit of the kinetic mixing parameter as
\begin{eqnarray}\label{epsilonupper}
\epsilon 
\,\ll\,
1.57 \times 10^{-6}
\bigg(\frac{\gD^{}}{0.3}\bigg)
\bigg(\frac{y^{}_\psi}{0.5}\bigg)^{\hs{-0.13cm}2}
\bigg(\frac{\upsilon_\phi}{50\,\tx{MeV}}\bigg)^{\hs{-0.13cm}2}
\bigg(\frac{m_\psi}{10\,\tx{GeV}}\bigg)^{\hs{-0.15cm}-2}
~.
\end{eqnarray}
Accordingly, with a proper choice of parameters, the DM can thermalize with the SM plasma in the early universe while evading the CMB constraint.\,\, Notice that with the upper limit value of the kinetic mixing parameter in Eq.\,\eqref{epsilonupper}, the DM-$e^-$ elastic scattering cross section is far below the current experimental sensitivities such as XENON1T~\cite{XENON:2019gfn}.\,\,Future prospective experiments using electron targets with improving sensitivities can test this model further.

The basic goal of this work was to show that the LSIDM is still a viable possibility and safe from the CMB bound.\,\,We have found a common parameter space that can provide the correct relic abundance satisfying region and the required self-interaction strength for DM.\,\,To explain the observed relic density, we have made the DM annihilation rate even stronger in contrast with the standard radiation-dominated universe that is very tightly constrained from CMB data in the region of parameter space we are interested in.\,\,We have built up a feasible model to evade the CMB bound through the decay of $Z'$ into the pairs of active neutrinos.\,\,Finally and most importantly, we have seen that even for larger values of $n$ and smaller values of $T_r$, the lower DM mass region (below ${\cal O}(1)\,\tx{GeV}$) remains below the light magenta and grey regions in Fig.\,\ref{fig:gDvsmN}.\,\,This means that if we can conclude that the DM is indeed self-interacting, moreover its mass lies in the sub-GeV range, one can then use these particle nature of DM to exclude the possibility of the existence of a fast-expanding universe.

\acknowledgments
This work is supported by KIAS Individual Grants under Grant No.\,PG081201 (SYH) and No.\,PG021403 (PK), and by National Research Foundation of Korea (NRF) Research Grant NRF-2019R1A2C3005009 (PK, DN). DN would also like to thank Debasish Borah and Satyabrata Mahapatra for the useful discussions.

\medskip
\textbf{\text{Note added --}}
During the finalization of this article, Ref.\,\cite{dutta2023selfinteracting} appeared on arXiv 
which also discusses the realization of self-interacting DM in the context of another non-standard expansion history of the universe.\,\,However, the author have not considered the CMB constraint which can significantly affect the allowed parameter space of the present setup, especially for the vector mediator case.

\bibliographystyle{JHEP}
\bibliography{ref}

\providecommand{\href}[2]{#2}\begingroup\raggedright\begin{thebibliography}{10}

\bibitem{Planck:2018vyg}
{\scshape Planck} collaboration, \emph{{Planck 2018 results. VI. Cosmological
  parameters}},
  \href{https://doi.org/10.1051/0004-6361/201833910}{\emph{Astron. Astrophys.}
  {\bfseries 641} (2020) A6}
  [\href{https://arxiv.org/abs/1807.06209}{{\ttfamily 1807.06209}}].

\bibitem{Feng:2010gw}
J.L.~Feng, \emph{{Dark Matter Candidates from Particle Physics and Methods of
  Detection}},
  \href{https://doi.org/10.1146/annurev-astro-082708-101659}{\emph{Ann. Rev.
  Astron. Astrophys.} {\bfseries 48} (2010) 495}
  [\href{https://arxiv.org/abs/1003.0904}{{\ttfamily 1003.0904}}].

\bibitem{Roszkowski:2017nbc}
L.~Roszkowski, E.M.~Sessolo and S.~Trojanowski, \emph{{WIMP dark matter
  candidates and searches\textemdash{}current status and future prospects}},
  \href{https://doi.org/10.1088/1361-6633/aab913}{\emph{Rept. Prog. Phys.}
  {\bfseries 81} (2018) 066201}
  [\href{https://arxiv.org/abs/1707.06277}{{\ttfamily 1707.06277}}].

\bibitem{Schumann:2019eaa}
M.~Schumann, \emph{{Direct Detection of WIMP Dark Matter: Concepts and
  Status}}, \href{https://doi.org/10.1088/1361-6471/ab2ea5}{\emph{J. Phys. G}
  {\bfseries 46} (2019) 103003}
  [\href{https://arxiv.org/abs/1903.03026}{{\ttfamily 1903.03026}}].

\bibitem{Lin:2019uvt}
T.~Lin, \emph{{Dark matter models and direct detection}},
  \href{https://doi.org/10.22323/1.333.0009}{\emph{PoS} {\bfseries 333} (2019)
  009} [\href{https://arxiv.org/abs/1904.07915}{{\ttfamily 1904.07915}}].

\bibitem{Arcadi:2017kky}
G.~Arcadi, M.~Dutra, P.~Ghosh, M.~Lindner, Y.~Mambrini, M.~Pierre et~al.,
  \emph{{The waning of the WIMP? A review of models, searches, and
  constraints}},
  \href{https://doi.org/10.1140/epjc/s10052-018-5662-y}{\emph{Eur. Phys. J. C}
  {\bfseries 78} (2018) 203}
  [\href{https://arxiv.org/abs/1703.07364}{{\ttfamily 1703.07364}}].

\bibitem{Spergel:1999mh}
D.N.~Spergel and P.J.~Steinhardt, \emph{{Observational evidence for
  selfinteracting cold dark matter}},
  \href{https://doi.org/10.1103/PhysRevLett.84.3760}{\emph{Phys. Rev. Lett.}
  {\bfseries 84} (2000) 3760}
  [\href{https://arxiv.org/abs/astro-ph/9909386}{{\ttfamily
  astro-ph/9909386}}].

\bibitem{Tulin:2017ara}
S.~Tulin and H.-B.~Yu, \emph{{Dark Matter Self-interactions and Small Scale
  Structure}}, \href{https://doi.org/10.1016/j.physrep.2017.11.004}{\emph{Phys.
  Rept.} {\bfseries 730} (2018) 1}
  [\href{https://arxiv.org/abs/1705.02358}{{\ttfamily 1705.02358}}].

\bibitem{Bullock:2017xww}
J.S.~Bullock and M.~Boylan-Kolchin, \emph{{Small-Scale Challenges to the
  $\Lambda$CDM Paradigm}},
  \href{https://doi.org/10.1146/annurev-astro-091916-055313}{\emph{Ann. Rev.
  Astron. Astrophys.} {\bfseries 55} (2017) 343}
  [\href{https://arxiv.org/abs/1707.04256}{{\ttfamily 1707.04256}}].

\bibitem{Ullio:2016kvy}
P.~Ullio and M.~Valli, \emph{{A critical reassessment of particle Dark Matter
  limits from dwarf satellites}},
  \href{https://doi.org/10.1088/1475-7516/2016/07/025}{\emph{JCAP} {\bfseries
  07} (2016) 025} [\href{https://arxiv.org/abs/1603.07721}{{\ttfamily
  1603.07721}}].

\bibitem{Blum:2016nrz}
K.~Blum, R.~Sato and T.R.~Slatyer, \emph{{Self-consistent Calculation of the
  Sommerfeld Enhancement}},
  \href{https://doi.org/10.1088/1475-7516/2016/06/021}{\emph{JCAP} {\bfseries
  06} (2016) 021} [\href{https://arxiv.org/abs/1603.01383}{{\ttfamily
  1603.01383}}].

\bibitem{Kamada:2016euw}
A.~Kamada, M.~Kaplinghat, A.B.~Pace and H.-B.~Yu, \emph{{How the
  Self-Interacting Dark Matter Model Explains the Diverse Galactic Rotation
  Curves}}, \href{https://doi.org/10.1103/PhysRevLett.119.111102}{\emph{Phys.
  Rev. Lett.} {\bfseries 119} (2017) 111102}
  [\href{https://arxiv.org/abs/1611.02716}{{\ttfamily 1611.02716}}].

\bibitem{Essig:2018pzq}
R.~Essig, S.D.~Mcdermott, H.-B.~Yu and Y.-M.~Zhong, \emph{{Constraining
  Dissipative Dark Matter Self-Interactions}},
  \href{https://doi.org/10.1103/PhysRevLett.123.121102}{\emph{Phys. Rev. Lett.}
  {\bfseries 123} (2019) 121102}
  [\href{https://arxiv.org/abs/1809.01144}{{\ttfamily 1809.01144}}].

\bibitem{Sagunski:2020spe}
L.~Sagunski, S.~Gad-Nasr, B.~Colquhoun, A.~Robertson and S.~Tulin,
  \emph{{Velocity-dependent Self-interacting Dark Matter from Groups and
  Clusters of Galaxies}},
  \href{https://doi.org/10.1088/1475-7516/2021/01/024}{\emph{JCAP} {\bfseries
  01} (2021) 024} [\href{https://arxiv.org/abs/2006.12515}{{\ttfamily
  2006.12515}}].

\bibitem{Colquhoun:2020adl}
B.~Colquhoun, S.~Heeba, F.~Kahlhoefer, L.~Sagunski and S.~Tulin,
  \emph{{Semiclassical regime for dark matter self-interactions}},
  \href{https://doi.org/10.1103/PhysRevD.103.035006}{\emph{Phys. Rev. D}
  {\bfseries 103} (2021) 035006}
  [\href{https://arxiv.org/abs/2011.04679}{{\ttfamily 2011.04679}}].

\bibitem{Eckert:2022qia}
D.~Eckert, S.~Ettori, A.~Robertson, R.~Massey, E.~Pointecouteau, D.~Harvey
  et~al., \emph{{Constraints on dark matter self-interaction from the internal
  density profiles of X-COP galaxy clusters}},
  \href{https://doi.org/10.1051/0004-6361/202243205}{\emph{Astron. Astrophys.}
  {\bfseries 666} (2022) A41}
  [\href{https://arxiv.org/abs/2205.01123}{{\ttfamily 2205.01123}}].

\bibitem{Silverman:2022bhs}
M.~Silverman, J.S.~Bullock, M.~Kaplinghat, V.H.~Robles and M.~Valli,
  \emph{{Motivations for a large self-interacting dark matter
  cross-section~from Milky Way satellites}},
  \href{https://doi.org/10.1093/mnras/stac3232}{\emph{Mon. Not. Roy. Astron.
  Soc.} {\bfseries 518} (2022) 2418}
  [\href{https://arxiv.org/abs/2203.10104}{{\ttfamily 2203.10104}}].

\bibitem{Binder:2022pmf}
T.~Binder, S.~Chakraborti, S.~Matsumoto and Y.~Watanabe, \emph{{A global
  analysis of resonance-enhanced light scalar dark matter}},
  \href{https://doi.org/10.1007/JHEP01(2023)106}{\emph{JHEP} {\bfseries 01}
  (2023) 106} [\href{https://arxiv.org/abs/2205.10149}{{\ttfamily
  2205.10149}}].

\bibitem{Girmohanta:2022dog}
S.~Girmohanta and R.~Shrock, \emph{{Cross section calculations in theories of
  self-interacting dark matter}},
  \href{https://doi.org/10.1103/PhysRevD.106.063013}{\emph{Phys. Rev. D}
  {\bfseries 106} (2022) 063013}
  [\href{https://arxiv.org/abs/2206.14395}{{\ttfamily 2206.14395}}].

\bibitem{Girmohanta:2022izb}
S.~Girmohanta and R.~Shrock, \emph{{Fitting a self-interacting dark matter
  model to data ranging from satellite galaxies to galaxy clusters}},
  \href{https://doi.org/10.1103/PhysRevD.107.063006}{\emph{Phys. Rev. D}
  {\bfseries 107} (2023) 063006}
  [\href{https://arxiv.org/abs/2210.01132}{{\ttfamily 2210.01132}}].

\bibitem{Buckley:2009in}
M.R.~Buckley and P.J.~Fox, \emph{{Dark Matter Self-Interactions and Light Force
  Carriers}}, \href{https://doi.org/10.1103/PhysRevD.81.083522}{\emph{Phys.
  Rev. D} {\bfseries 81} (2010) 083522}
  [\href{https://arxiv.org/abs/0911.3898}{{\ttfamily 0911.3898}}].

\bibitem{Feng:2009hw}
J.L.~Feng, M.~Kaplinghat and H.-B.~Yu, \emph{{Halo Shape and Relic Density
  Exclusions of Sommerfeld-Enhanced Dark Matter Explanations of Cosmic Ray
  Excesses}}, \href{https://doi.org/10.1103/PhysRevLett.104.151301}{\emph{Phys.
  Rev. Lett.} {\bfseries 104} (2010) 151301}
  [\href{https://arxiv.org/abs/0911.0422}{{\ttfamily 0911.0422}}].

\bibitem{Feng:2009mn}
J.L.~Feng, M.~Kaplinghat, H.~Tu and H.-B.~Yu, \emph{{Hidden Charged Dark
  Matter}}, \href{https://doi.org/10.1088/1475-7516/2009/07/004}{\emph{JCAP}
  {\bfseries 07} (2009) 004} [\href{https://arxiv.org/abs/0905.3039}{{\ttfamily
  0905.3039}}].

\bibitem{Loeb:2010gj}
A.~Loeb and N.~Weiner, \emph{{Cores in Dwarf Galaxies from Dark Matter with a
  Yukawa Potential}},
  \href{https://doi.org/10.1103/PhysRevLett.106.171302}{\emph{Phys. Rev. Lett.}
  {\bfseries 106} (2011) 171302}
  [\href{https://arxiv.org/abs/1011.6374}{{\ttfamily 1011.6374}}].

\bibitem{Vogelsberger:2012ku}
M.~Vogelsberger, J.~Zavala and A.~Loeb, \emph{{Subhaloes in Self-Interacting
  Galactic Dark Matter Haloes}},
  \href{https://doi.org/10.1111/j.1365-2966.2012.21182.x}{\emph{Mon. Not. Roy.
  Astron. Soc.} {\bfseries 423} (2012) 3740}
  [\href{https://arxiv.org/abs/1201.5892}{{\ttfamily 1201.5892}}].

\bibitem{Borah:2021pet}
D.~Borah, M.~Dutta, S.~Mahapatra and N.~Sahu, \emph{{Self-interacting dark
  matter via right handed neutrino portal}},
  \href{https://doi.org/10.1103/PhysRevD.105.015004}{\emph{Phys. Rev. D}
  {\bfseries 105} (2022) 015004}
  [\href{https://arxiv.org/abs/2110.00021}{{\ttfamily 2110.00021}}].

\bibitem{Borah:2022ask}
D.~Borah, S.~Mahapatra and N.~Sahu, \emph{{Minimal realisation of light thermal
  self-interacting dark matter}},
  \href{https://arxiv.org/abs/2211.15703}{{\ttfamily 2211.15703}}.

\bibitem{Slatyer:2015jla}
T.R.~Slatyer, \emph{{Indirect dark matter signatures in the cosmic dark ages.
  I. Generalizing the bound on s-wave dark matter annihilation from Planck
  results}}, \href{https://doi.org/10.1103/PhysRevD.93.023527}{\emph{Phys. Rev.
  D} {\bfseries 93} (2016) 023527}
  [\href{https://arxiv.org/abs/1506.03811}{{\ttfamily 1506.03811}}].

\bibitem{Galli:2011rz}
S.~Galli, F.~Iocco, G.~Bertone and A.~Melchiorri, \emph{{Updated CMB
  constraints on Dark Matter annihilation cross-sections}},
  \href{https://doi.org/10.1103/PhysRevD.84.027302}{\emph{Phys. Rev. D}
  {\bfseries 84} (2011) 027302}
  [\href{https://arxiv.org/abs/1106.1528}{{\ttfamily 1106.1528}}].

\bibitem{Giesen:2012rp}
G.~Giesen, J.~Lesgourgues, B.~Audren and Y.~Ali-Haimoud, \emph{{CMB photons
  shedding light on dark matter}},
  \href{https://doi.org/10.1088/1475-7516/2012/12/008}{\emph{JCAP} {\bfseries
  12} (2012) 008} [\href{https://arxiv.org/abs/1209.0247}{{\ttfamily
  1209.0247}}].

\bibitem{Griest:1990kh}
K.~Griest and D.~Seckel, \emph{{Three exceptions in the calculation of relic
  abundances}}, \href{https://doi.org/10.1103/PhysRevD.43.3191}{\emph{Phys.
  Rev. D} {\bfseries 43} (1991) 3191}.

\bibitem{Kaplan:2009ag}
D.E.~Kaplan, M.A.~Luty and K.M.~Zurek, \emph{{Asymmetric Dark Matter}},
  \href{https://doi.org/10.1103/PhysRevD.79.115016}{\emph{Phys. Rev. D}
  {\bfseries 79} (2009) 115016}
  [\href{https://arxiv.org/abs/0901.4117}{{\ttfamily 0901.4117}}].

\bibitem{Iminniyaz:2011yp}
H.~Iminniyaz, M.~Drees and X.~Chen, \emph{{Relic Abundance of Asymmetric Dark
  Matter}}, \href{https://doi.org/10.1088/1475-7516/2011/07/003}{\emph{JCAP}
  {\bfseries 07} (2011) 003} [\href{https://arxiv.org/abs/1104.5548}{{\ttfamily
  1104.5548}}].

\bibitem{Graesser:2011wi}
M.L.~Graesser, I.M.~Shoemaker and L.~Vecchi, \emph{{Asymmetric WIMP dark
  matter}}, \href{https://doi.org/10.1007/JHEP10(2011)110}{\emph{JHEP}
  {\bfseries 10} (2011) 110} [\href{https://arxiv.org/abs/1103.2771}{{\ttfamily
  1103.2771}}].

\bibitem{Ho:2022tbw}
S.-Y.~Ho, \emph{{An asymmetric SIMP dark matter model}},
  \href{https://doi.org/10.1007/JHEP10(2022)182}{\emph{JHEP} {\bfseries 10}
  (2022) 182} [\href{https://arxiv.org/abs/2207.13373}{{\ttfamily
  2207.13373}}].

\bibitem{Chen:2023rrl}
Z.~Chen, K.~Ye and M.~Zhang, \emph{{Asymmetric dark matter with a spontaneously
  broken U(1)': Self-interaction and gravitational waves}},
  \href{https://doi.org/10.1103/PhysRevD.107.095027}{\emph{Phys. Rev. D}
  {\bfseries 107} (2023) 095027}
  [\href{https://arxiv.org/abs/2303.11820}{{\ttfamily 2303.11820}}].

\bibitem{DAgnolo:2015ujb}
R.T.~D'Agnolo and J.T.~Ruderman, \emph{{Light Dark Matter from Forbidden
  Channels}}, \href{https://doi.org/10.1103/PhysRevLett.115.061301}{\emph{Phys.
  Rev. Lett.} {\bfseries 115} (2015) 061301}
  [\href{https://arxiv.org/abs/1505.07107}{{\ttfamily 1505.07107}}].

\bibitem{DAgnolo:2020mpt}
R.T.~D'Agnolo, D.~Liu, J.T.~Ruderman and P.-J.~Wang, \emph{{Forbidden dark
  matter annihilations into Standard Model particles}},
  \href{https://doi.org/10.1007/JHEP06(2021)103}{\emph{JHEP} {\bfseries 06}
  (2021) 103} [\href{https://arxiv.org/abs/2012.11766}{{\ttfamily
  2012.11766}}].

\bibitem{Herms:2022nhd}
J.~Herms, S.~Jana, V.P.~K. and S.~Saad, \emph{{Minimal Realization of Light
  Thermal Dark Matter}},
  \href{https://doi.org/10.1103/PhysRevLett.129.091803}{\emph{Phys. Rev. Lett.}
  {\bfseries 129} (2022) 091803}
  [\href{https://arxiv.org/abs/2203.05579}{{\ttfamily 2203.05579}}].

\bibitem{Hochberg:2014dra}
Y.~Hochberg, E.~Kuflik, T.~Volansky and J.G.~Wacker, \emph{{Mechanism for
  Thermal Relic Dark Matter of Strongly Interacting Massive Particles}},
  \href{https://doi.org/10.1103/PhysRevLett.113.171301}{\emph{Phys. Rev. Lett.}
  {\bfseries 113} (2014) 171301}
  [\href{https://arxiv.org/abs/1402.5143}{{\ttfamily 1402.5143}}].

\bibitem{Profumo:2017obk}
S.~Profumo, F.S.~Queiroz, J.~Silk and C.~Siqueira, \emph{{Searching for
  Secluded Dark Matter with H.E.S.S., Fermi-LAT, and Planck}},
  \href{https://doi.org/10.1088/1475-7516/2018/03/010}{\emph{JCAP} {\bfseries
  03} (2018) 010} [\href{https://arxiv.org/abs/1711.03133}{{\ttfamily
  1711.03133}}].

\bibitem{Kawasaki:2000en}
M.~Kawasaki, K.~Kohri and N.~Sugiyama, \emph{{MeV scale reheating temperature
  and thermalization of neutrino background}},
  \href{https://doi.org/10.1103/PhysRevD.62.023506}{\emph{Phys. Rev. D}
  {\bfseries 62} (2000) 023506}
  [\href{https://arxiv.org/abs/astro-ph/0002127}{{\ttfamily
  astro-ph/0002127}}].

\bibitem{Ichikawa:2005vw}
K.~Ichikawa, M.~Kawasaki and F.~Takahashi, \emph{{The Oscillation effects on
  thermalization of the neutrinos in the Universe with low reheating
  temperature}}, \href{https://doi.org/10.1103/PhysRevD.72.043522}{\emph{Phys.
  Rev. D} {\bfseries 72} (2005) 043522}
  [\href{https://arxiv.org/abs/astro-ph/0505395}{{\ttfamily
  astro-ph/0505395}}].

\bibitem{Chung:1998rq}
D.J.H.~Chung, E.W.~Kolb and A.~Riotto, \emph{{Production of massive particles
  during reheating}},
  \href{https://doi.org/10.1103/PhysRevD.60.063504}{\emph{Phys. Rev. D}
  {\bfseries 60} (1999) 063504}
  [\href{https://arxiv.org/abs/hep-ph/9809453}{{\ttfamily hep-ph/9809453}}].

\bibitem{Moroi:1999zb}
T.~Moroi and L.~Randall, \emph{{Wino cold dark matter from anomaly mediated
  SUSY breaking}},
  \href{https://doi.org/10.1016/S0550-3213(99)00748-8}{\emph{Nucl. Phys. B}
  {\bfseries 570} (2000) 455}
  [\href{https://arxiv.org/abs/hep-ph/9906527}{{\ttfamily hep-ph/9906527}}].

\bibitem{Giudice:2000ex}
G.F.~Giudice, E.W.~Kolb and A.~Riotto, \emph{{Largest temperature of the
  radiation era and its cosmological implications}},
  \href{https://doi.org/10.1103/PhysRevD.64.023508}{\emph{Phys. Rev. D}
  {\bfseries 64} (2001) 023508}
  [\href{https://arxiv.org/abs/hep-ph/0005123}{{\ttfamily hep-ph/0005123}}].

\bibitem{Allahverdi:2002nb}
R.~Allahverdi and M.~Drees, \emph{{Production of massive stable particles in
  inflaton decay}},
  \href{https://doi.org/10.1103/PhysRevLett.89.091302}{\emph{Phys. Rev. Lett.}
  {\bfseries 89} (2002) 091302}
  [\href{https://arxiv.org/abs/hep-ph/0203118}{{\ttfamily hep-ph/0203118}}].

\bibitem{Allahverdi:2002pu}
R.~Allahverdi and M.~Drees, \emph{{Thermalization after inflation and
  production of massive stable particles}},
  \href{https://doi.org/10.1103/PhysRevD.66.063513}{\emph{Phys. Rev. D}
  {\bfseries 66} (2002) 063513}
  [\href{https://arxiv.org/abs/hep-ph/0205246}{{\ttfamily hep-ph/0205246}}].

\bibitem{Acharya:2009zt}
B.S.~Acharya, G.~Kane, S.~Watson and P.~Kumar, \emph{{A Non-thermal WIMP
  Miracle}}, \href{https://doi.org/10.1103/PhysRevD.80.083529}{\emph{Phys. Rev.
  D} {\bfseries 80} (2009) 083529}
  [\href{https://arxiv.org/abs/0908.2430}{{\ttfamily 0908.2430}}].

\bibitem{Allahverdi:2010xz}
R.~Allahverdi, R.~Brandenberger, F.-Y.~Cyr-Racine and A.~Mazumdar,
  \emph{{Reheating in Inflationary Cosmology: Theory and Applications}},
  \href{https://doi.org/10.1146/annurev.nucl.012809.104511}{\emph{Ann. Rev.
  Nucl. Part. Sci.} {\bfseries 60} (2010) 27}
  [\href{https://arxiv.org/abs/1001.2600}{{\ttfamily 1001.2600}}].

\bibitem{Monteux:2015qqa}
A.~Monteux and C.S.~Shin, \emph{{Thermal Goldstino Production with Low
  Reheating Temperatures}},
  \href{https://doi.org/10.1103/PhysRevD.92.035002}{\emph{Phys. Rev. D}
  {\bfseries 92} (2015) 035002}
  [\href{https://arxiv.org/abs/1505.03149}{{\ttfamily 1505.03149}}].

\bibitem{Davoudiasl:2015vba}
H.~Davoudiasl, D.~Hooper and S.D.~McDermott, \emph{{Inflatable Dark Matter}},
  \href{https://doi.org/10.1103/PhysRevLett.116.031303}{\emph{Phys. Rev. Lett.}
  {\bfseries 116} (2016) 031303}
  [\href{https://arxiv.org/abs/1507.08660}{{\ttfamily 1507.08660}}].

\bibitem{Berlin:2016vnh}
A.~Berlin, D.~Hooper and G.~Krnjaic, \emph{{PeV-Scale Dark Matter as a Thermal
  Relic of a Decoupled Sector}},
  \href{https://doi.org/10.1016/j.physletb.2016.06.037}{\emph{Phys. Lett. B}
  {\bfseries 760} (2016) 106}
  [\href{https://arxiv.org/abs/1602.08490}{{\ttfamily 1602.08490}}].

\bibitem{Tenkanen:2016jic}
T.~Tenkanen and V.~Vaskonen, \emph{{Reheating the Standard Model from a hidden
  sector}}, \href{https://doi.org/10.1103/PhysRevD.94.083516}{\emph{Phys. Rev.
  D} {\bfseries 94} (2016) 083516}
  [\href{https://arxiv.org/abs/1606.00192}{{\ttfamily 1606.00192}}].

\bibitem{Berlin:2016gtr}
A.~Berlin, D.~Hooper and G.~Krnjaic, \emph{{Thermal Dark Matter From A Highly
  Decoupled Sector}},
  \href{https://doi.org/10.1103/PhysRevD.94.095019}{\emph{Phys. Rev. D}
  {\bfseries 94} (2016) 095019}
  [\href{https://arxiv.org/abs/1609.02555}{{\ttfamily 1609.02555}}].

\bibitem{DEramo:2017gpl}
F.~D'Eramo, N.~Fernandez and S.~Profumo, \emph{{When the Universe Expands Too
  Fast: Relentless Dark Matter}},
  \href{https://doi.org/10.1088/1475-7516/2017/05/012}{\emph{JCAP} {\bfseries
  05} (2017) 012} [\href{https://arxiv.org/abs/1703.04793}{{\ttfamily
  1703.04793}}].

\bibitem{DEramo:2017ecx}
F.~D'Eramo, N.~Fernandez and S.~Profumo, \emph{{Dark Matter Freeze-in
  Production in Fast-Expanding Universes}},
  \href{https://doi.org/10.1088/1475-7516/2018/02/046}{\emph{JCAP} {\bfseries
  02} (2018) 046} [\href{https://arxiv.org/abs/1712.07453}{{\ttfamily
  1712.07453}}].

\bibitem{Bernal:2019uqr}
N.~Bernal, X.~Chu, S.~Kulkarni and J.~Pradler, \emph{{Self-interacting dark
  matter without prejudice}},
  \href{https://doi.org/10.1103/PhysRevD.101.055044}{\emph{Phys. Rev. D}
  {\bfseries 101} (2020) 055044}
  [\href{https://arxiv.org/abs/1912.06681}{{\ttfamily 1912.06681}}].

\bibitem{Allahverdi:2020bys}
R.~Allahverdi et~al., \emph{{The First Three Seconds: a Review of Possible
  Expansion Histories of the Early Universe}},
  \href{https://arxiv.org/abs/2006.16182}{{\ttfamily 2006.16182}}.

\bibitem{Ghosh:2021wrk}
P.~Ghosh, P.~Konar, A.K.~Saha and S.~Show, \emph{{Self-interacting freeze-in
  dark matter in a singlet doublet scenario}},
  \href{https://doi.org/10.1088/1475-7516/2022/10/017}{\emph{JCAP} {\bfseries
  10} (2022) 017} [\href{https://arxiv.org/abs/2112.09057}{{\ttfamily
  2112.09057}}].

\bibitem{Biondini:2023hek}
S.~Biondini, \emph{{Interplay between improved interaction rates and modified
  cosmological histories for dark matter}},  9, 2023
  [\href{https://arxiv.org/abs/2309.00323}{{\ttfamily 2309.00323}}].

\bibitem{Scherrer:2022nnz}
R.J.~Scherrer, \emph{{How slowly can the early Universe expand?}},
  \href{https://doi.org/10.1103/PhysRevD.106.103516}{\emph{Phys. Rev. D}
  {\bfseries 106} (2022) 103516}
  [\href{https://arxiv.org/abs/2209.03421}{{\ttfamily 2209.03421}}].

\bibitem{Biswas:2022fga}
A.~Biswas, D.K.~Ghosh and D.~Nanda, \emph{{Concealing Dirac neutrinos from
  cosmic microwave background}},
  \href{https://doi.org/10.1088/1475-7516/2022/10/006}{\emph{JCAP} {\bfseries
  10} (2022) 006} [\href{https://arxiv.org/abs/2206.13710}{{\ttfamily
  2206.13710}}].

\bibitem{Turner:1983he}
M.S.~Turner, \emph{{Coherent Scalar Field Oscillations in an Expanding
  Universe}}, \href{https://doi.org/10.1103/PhysRevD.28.1243}{\emph{Phys. Rev.
  D} {\bfseries 28} (1983) 1243}.

\bibitem{Ratra:1987rm}
B.~Ratra and P.J.E.~Peebles, \emph{{Cosmological Consequences of a Rolling
  Homogeneous Scalar Field}},
  \href{https://doi.org/10.1103/PhysRevD.37.3406}{\emph{Phys. Rev. D}
  {\bfseries 37} (1988) 3406}.

\bibitem{Joyce:1996cp}
M.~Joyce, \emph{{Electroweak Baryogenesis and the Expansion Rate of the
  Universe}}, \href{https://doi.org/10.1103/PhysRevD.55.1875}{\emph{Phys. Rev.
  D} {\bfseries 55} (1997) 1875}
  [\href{https://arxiv.org/abs/hep-ph/9606223}{{\ttfamily hep-ph/9606223}}].

\bibitem{Copeland:1997et}
E.J.~Copeland, A.R.~Liddle and D.~Wands, \emph{{Exponential potentials and
  cosmological scaling solutions}},
  \href{https://doi.org/10.1103/PhysRevD.57.4686}{\emph{Phys. Rev. D}
  {\bfseries 57} (1998) 4686}
  [\href{https://arxiv.org/abs/gr-qc/9711068}{{\ttfamily gr-qc/9711068}}].

\bibitem{Ferreira:1997hj}
P.G.~Ferreira and M.~Joyce, \emph{{Cosmology with a primordial scaling field}},
  \href{https://doi.org/10.1103/PhysRevD.58.023503}{\emph{Phys. Rev. D}
  {\bfseries 58} (1998) 023503}
  [\href{https://arxiv.org/abs/astro-ph/9711102}{{\ttfamily
  astro-ph/9711102}}].

\bibitem{Ferreira:1997au}
P.G.~Ferreira and M.~Joyce, \emph{{Structure formation with a selftuning scalar
  field}}, \href{https://doi.org/10.1103/PhysRevLett.79.4740}{\emph{Phys. Rev.
  Lett.} {\bfseries 79} (1997) 4740}
  [\href{https://arxiv.org/abs/astro-ph/9707286}{{\ttfamily
  astro-ph/9707286}}].

\bibitem{Kallosh:2013hoa}
R.~Kallosh and A.~Linde, \emph{{Universality Class in Conformal Inflation}},
  \href{https://doi.org/10.1088/1475-7516/2013/07/002}{\emph{JCAP} {\bfseries
  07} (2013) 002} [\href{https://arxiv.org/abs/1306.5220}{{\ttfamily
  1306.5220}}].

\bibitem{Kallosh:2013yoa}
R.~Kallosh, A.~Linde and D.~Roest, \emph{{Superconformal Inflationary
  $\alpha$-Attractors}},
  \href{https://doi.org/10.1007/JHEP11(2013)198}{\emph{JHEP} {\bfseries 11}
  (2013) 198} [\href{https://arxiv.org/abs/1311.0472}{{\ttfamily 1311.0472}}].

\bibitem{Wands:1993zm}
D.~Wands, E.J.~Copeland and A.R.~Liddle, \emph{{Exponential potentials, scaling
  solutions and inflation}},  in \emph{{16th Texas Symposium on Relativistic
  Astrophysics and 3rd Particles, Strings, and Cosmology Symposium}},
  pp.~0647--652, 3, 1993.

\bibitem{Caldwell:1997ii}
R.R.~Caldwell, R.~Dave and P.J.~Steinhardt, \emph{{Cosmological imprint of an
  energy component with general equation of state}},
  \href{https://doi.org/10.1103/PhysRevLett.80.1582}{\emph{Phys. Rev. Lett.}
  {\bfseries 80} (1998) 1582}
  [\href{https://arxiv.org/abs/astro-ph/9708069}{{\ttfamily
  astro-ph/9708069}}].

\bibitem{Wetterich:1994bg}
C.~Wetterich, \emph{{The Cosmon model for an asymptotically vanishing time
  dependent cosmological 'constant'}}, {\emph{Astron. Astrophys.} {\bfseries
  301} (1995) 321} [\href{https://arxiv.org/abs/hep-th/9408025}{{\ttfamily
  hep-th/9408025}}].

\bibitem{Sahni:1999gb}
V.~Sahni and A.A.~Starobinsky, \emph{{The Case for a positive cosmological
  Lambda term}}, \href{https://doi.org/10.1142/S0218271800000542}{\emph{Int. J.
  Mod. Phys. D} {\bfseries 9} (2000) 373}
  [\href{https://arxiv.org/abs/astro-ph/9904398}{{\ttfamily
  astro-ph/9904398}}].

\bibitem{Saikawa:2018rcs}
K.~Saikawa and S.~Shirai, \emph{{Primordial gravitational waves, precisely: The
  role of thermodynamics in the Standard Model}},
  \href{https://doi.org/10.1088/1475-7516/2018/05/035}{\emph{JCAP} {\bfseries
  05} (2018) 035} [\href{https://arxiv.org/abs/1803.01038}{{\ttfamily
  1803.01038}}].

\bibitem{Ko:2012hd}
P.~Ko, Y.~Omura and C.~Yu, \emph{{A Resolution of the Flavor Problem of Two
  Higgs Doublet Models with an Extra $U(1)_H$ Symmetry for Higgs Flavor}},
  \href{https://doi.org/10.1016/j.physletb.2012.09.019}{\emph{Phys. Lett. B}
  {\bfseries 717} (2012) 202}
  [\href{https://arxiv.org/abs/1204.4588}{{\ttfamily 1204.4588}}].

\bibitem{Pospelov:2007mp}
M.~Pospelov, A.~Ritz and M.B.~Voloshin, \emph{{Secluded WIMP Dark Matter}},
  \href{https://doi.org/10.1016/j.physletb.2008.02.052}{\emph{Phys. Lett. B}
  {\bfseries 662} (2008) 53} [\href{https://arxiv.org/abs/0711.4866}{{\ttfamily
  0711.4866}}].

\bibitem{Escudero:2019gzq}
M.~Escudero, D.~Hooper, G.~Krnjaic and M.~Pierre, \emph{{Cosmology with A Very
  Light L$_{\mu}$ \ensuremath{-} L$_{\tau}$ Gauge Boson}},
  \href{https://doi.org/10.1007/JHEP03(2019)071}{\emph{JHEP} {\bfseries 03}
  (2019) 071} [\href{https://arxiv.org/abs/1901.02010}{{\ttfamily
  1901.02010}}].

\bibitem{Berbig:2020wve}
M.~Berbig, S.~Jana and A.~Trautner, \emph{{The Hubble tension and a
  renormalizable model of gauged neutrino self-interactions}},
  \href{https://doi.org/10.1103/PhysRevD.102.115008}{\emph{Phys. Rev. D}
  {\bfseries 102} (2020) 115008}
  [\href{https://arxiv.org/abs/2004.13039}{{\ttfamily 2004.13039}}].

\bibitem{Farzan:2016wym}
Y.~Farzan and J.~Heeck, \emph{{Neutrinophilic nonstandard interactions}},
  \href{https://doi.org/10.1103/PhysRevD.94.053010}{\emph{Phys. Rev. D}
  {\bfseries 94} (2016) 053010}
  [\href{https://arxiv.org/abs/1607.07616}{{\ttfamily 1607.07616}}].

\bibitem{Denton:2018dqq}
P.B.~Denton, Y.~Farzan and I.M.~Shoemaker, \emph{{Activating the fourth
  neutrino of the 3+1 scheme}},
  \href{https://doi.org/10.1103/PhysRevD.99.035003}{\emph{Phys. Rev. D}
  {\bfseries 99} (2019) 035003}
  [\href{https://arxiv.org/abs/1811.01310}{{\ttfamily 1811.01310}}].

\bibitem{Kayser:2002qs}
B.~Kayser, \emph{{Neutrino Mass, Mixing, and Flavor Change}},
  \href{https://arxiv.org/abs/hep-ph/0211134}{{\ttfamily hep-ph/0211134}}.

\bibitem{Bhattacharya:2019mmy}
S.~Bhattacharya, P.~Ghosh and S.~Verma, \emph{{SIMPler realisation of Scalar
  Dark Matter}},
  \href{https://doi.org/10.1088/1475-7516/2020/01/040}{\emph{JCAP} {\bfseries
  01} (2020) 040} [\href{https://arxiv.org/abs/1904.07562}{{\ttfamily
  1904.07562}}].

\bibitem{Tulin:2013teo}
S.~Tulin, H.-B.~Yu and K.M.~Zurek, \emph{{Beyond Collisionless Dark Matter:
  Particle Physics Dynamics for Dark Matter Halo Structure}},
  \href{https://doi.org/10.1103/PhysRevD.87.115007}{\emph{Phys. Rev. D}
  {\bfseries 87} (2013) 115007}
  [\href{https://arxiv.org/abs/1302.3898}{{\ttfamily 1302.3898}}].

\bibitem{Tulin:2012wi}
S.~Tulin, H.-B.~Yu and K.M.~Zurek, \emph{{Resonant Dark Forces and Small Scale
  Structure}},
  \href{https://doi.org/10.1103/PhysRevLett.110.111301}{\emph{Phys. Rev. Lett.}
  {\bfseries 110} (2013) 111301}
  [\href{https://arxiv.org/abs/1210.0900}{{\ttfamily 1210.0900}}].

\bibitem{Khrapak:2003kjw}
S.A.~Khrapak, A.V.~Ivlev, G.E.~Morfill and S.K.~Zhdanov, \emph{{Scattering in
  the Attractive Yukawa Potential in the Limit of Strong Interaction}},
  \href{https://doi.org/10.1103/PhysRevLett.90.225002}{\emph{Phys. Rev. Lett.}
  {\bfseries 90} (2003) 225002}.

\bibitem{Kaplinghat:2015aga}
M.~Kaplinghat, S.~Tulin and H.-B.~Yu, \emph{{Dark Matter Halos as Particle
  Colliders: Unified Solution to Small-Scale Structure Puzzles from Dwarfs to
  Clusters}}, \href{https://doi.org/10.1103/PhysRevLett.116.041302}{\emph{Phys.
  Rev. Lett.} {\bfseries 116} (2016) 041302}
  [\href{https://arxiv.org/abs/1508.03339}{{\ttfamily 1508.03339}}].

\bibitem{Grimus:2007if}
W.~Grimus, L.~Lavoura, O.M.~Ogreid and P.~Osland, \emph{{A Precision constraint
  on multi-Higgs-doublet models}},
  \href{https://doi.org/10.1088/0954-3899/35/7/075001}{\emph{J. Phys. G}
  {\bfseries 35} (2008) 075001}
  [\href{https://arxiv.org/abs/0711.4022}{{\ttfamily 0711.4022}}].

\bibitem{Grimus:2008nb}
W.~Grimus, L.~Lavoura, O.M.~Ogreid and P.~Osland, \emph{{The Oblique parameters
  in multi-Higgs-doublet models}},
  \href{https://doi.org/10.1016/j.nuclphysb.2008.04.019}{\emph{Nucl. Phys. B}
  {\bfseries 801} (2008) 81} [\href{https://arxiv.org/abs/0802.4353}{{\ttfamily
  0802.4353}}].

\bibitem{ParticleDataGroup:2022pth}
{\scshape Particle Data Group} collaboration, \emph{{Review of Particle
  Physics}}, \href{https://doi.org/10.1093/ptep/ptac097}{\emph{PTEP} {\bfseries
  2022} (2022) 083C01}.

\bibitem{CMS:2021kom}
{\scshape CMS} collaboration, \emph{{Measurements of Higgs boson production
  cross sections and couplings in the diphoton decay channel at $
  \sqrt{\mathrm{s}} $ = 13 TeV}},
  \href{https://doi.org/10.1007/JHEP07(2021)027}{\emph{JHEP} {\bfseries 07}
  (2021) 027} [\href{https://arxiv.org/abs/2103.06956}{{\ttfamily
  2103.06956}}].

\bibitem{Djouadi:2005gi}
A.~Djouadi, \emph{{The Anatomy of electro-weak symmetry breaking. I: The Higgs
  boson in the standard model}},
  \href{https://doi.org/10.1016/j.physrep.2007.10.004}{\emph{Phys. Rept.}
  {\bfseries 457} (2008) 1}
  [\href{https://arxiv.org/abs/hep-ph/0503172}{{\ttfamily hep-ph/0503172}}].

\bibitem{Gondolo:2012vh}
P.~Gondolo, J.~Hisano and K.~Kadota, \emph{{The Effect of quark interactions on
  dark matter kinetic decoupling and the mass of the smallest dark halos}},
  \href{https://doi.org/10.1103/PhysRevD.86.083523}{\emph{Phys. Rev. D}
  {\bfseries 86} (2012) 083523}
  [\href{https://arxiv.org/abs/1205.1914}{{\ttfamily 1205.1914}}].

\bibitem{Ho:2022erb}
S.-Y.~Ho, P.~Ko and C.-T.~Lu, \emph{{Scalar and fermion two-component SIMP dark
  matter with an accidental $\mathbb Z_{4}$ symmetry}},
  \href{https://doi.org/10.1007/JHEP03(2022)005}{\emph{JHEP} {\bfseries 03}
  (2022) 005} [\href{https://arxiv.org/abs/2201.06856}{{\ttfamily
  2201.06856}}].

\bibitem{XENON:2019gfn}
{\scshape XENON} collaboration, \emph{{Light Dark Matter Search with Ionization
  Signals in XENON1T}},
  \href{https://doi.org/10.1103/PhysRevLett.123.251801}{\emph{Phys. Rev. Lett.}
  {\bfseries 123} (2019) 251801}
  [\href{https://arxiv.org/abs/1907.11485}{{\ttfamily 1907.11485}}].

\bibitem{dutta2023selfinteracting}
M.~Dutta, \emph{Self-interacting dark matter in non-standard cosmology},
  \href{https://arxiv.org/abs/2310.03909}{{\ttfamily 2310.03909}}.

\end{thebibliography}\endgroup

\end{document}